\documentstyle[psfig]{mn2e}

\def\gs{\mathrel{\raise0.35ex\hbox{$\scriptstyle >$}\kern-0.6em
\lower0.40ex\hbox{{$\scriptstyle \sim$}}}}
\def\ls{\mathrel{\raise0.35ex\hbox{$\scriptstyle <$}\kern-0.6em
\lower0.40ex\hbox{{$\scriptstyle \sim$}}}}
\def\ls{\mathrel{\hbox{\rlap{\hbox{\lower4pt\hbox{$\sim$}}}\hbox{$<$}}}}
\def\gs{\mathrel{\hbox{\rlap{\hbox{\lower4pt\hbox{$\sim$}}}\hbox{$>$}}}}

\title[Colour gradients across intermediate $L_X$ galaxy clusters]
      {Red sequence modal colour gradients across intermediate X-ray luminosity galaxy clusters}
\author[Jensen and Pimbblet]
       {Peter C.\ Jensen$^{1}$\thanks{email:pjensen@swin.edu.au} and Kevin A.\ Pimbblet$^{2,3}$\thanks{email:kevin.pimbblet@monash.edu} 
        \vspace*{1mm}\\
$^{1}$Centre for Astrophysics and Supercomputing, 
Swinburne University of Technology, Hawthorn, Victoria 3122, Australia\\
$^{2}$School of Physics, Monash University, Clayton, Victoria 3800, Australia\\
$^{3}$Monash Centre for Astrophysics (MoCA), Monash University, Clayton, Victoria 3800, Australia}

\date{\fbox{\sc Draft: \today\ --- Do Not Distribute}}

\pagerange{000--000}

\begin{document}

\maketitle

\begin{abstract}
We assemble a sample of 45 intermediate X-ray luminosity galaxy clusters
($0.7\times10^{44} < L_X < 4\times10^{44}$ ergs$^{-1}$) at low
redshifts ($0.03<z<0.16$) using SDSS data to conduct a comprehensive
investigation into the photometric variation of 
red sequence modal galaxy colours with environment.
The clusters span a range of Bautz-Morgan types and evolutionary 
stages and are shown to be representative of the global underlying intermediate
$L_X$ cluster sample.
We define cluster membership using SDSS spectroscopic data 
and characterize the clusters by deriving 
new recession velocities, velocity dispersions and other parameters for each.  
We construct colour-magnitude diagrams for each of these clusters and 
obtain the position of the red sequence using a robust
line fitting algorithm with a Lorentzian merit function.
In doing so, we describe a population of discordant points on the colour-magnitude
plane which are the result of photometric blending, dust and other causes.
By fitting the clusters with Schechter functions to derive $M^{\star}$
values in each SDSS passband, we combine the red sequence of the
galaxy clusters together to form a composite sample.
We detail how the modal colour value of the red sequence
varies with radius from the centre of this composite cluster 
and local galaxy density for all SDSS colours.
In agreement with previous studies, these colours are shown
to systematically move blueward with increasing distance
from the cluster centres, or equivalently lower local galaxy density,
whilst the width of the red sequence increases.  This supports the
idea that the galaxies at the outskirts of these clusters have
younger luminosity-weighted ages than those at the core
indicating their star formation has been quenched more recently
than in the core.
A comparison of our derived gradients in $(g-r)$ 
(explicitly: $d(g-r)/d log(r_p) = -0.031 \pm 0.003$ and 
$d(g-r)/d log(\Sigma) = 0.012 \pm 0.002$)
with earlier works tentatively suggests that these gradients 
vary redshift which would
reflect the hierarchical build-up of the red sequence over time.

\end{abstract}


\section{Introduction}
A lot of effort has been devoted to understanding how galaxies evolve in clusters 
over recent years.
Most of the attention has been placed on the highest-mass clusters,
largely because they are the 
easiest ones to detect to high-redshift, and therefore can be contrasted
to lower redshift analogues
(e.g.\ Yee et al.\ 1996; Balogh et al.\ 1999; Fasano et al.\ 2000;
Ebeling et al.\ 2001; Fairley et al.\ 2002; Poggianti et al.\ 2004;
Wake et al.\ 2005; Pimbblet et al.\ 2006; Poggianti et al.\ 2009; Valentinuzzi et al.\ 2011).
With a few notable exceptions 
(e.g.\ Andreon et al.\ 2004; the WINGS survey of Fasano et al.\ 2006), 
intermediate and low-mass clusters ($L_X \sim 1 \times 10^{44}$ erg s$^{-1}$) 
are a relatively unexplored region of parameter space and thus merit attention.

\begin{figure}
\centerline{\psfig{file=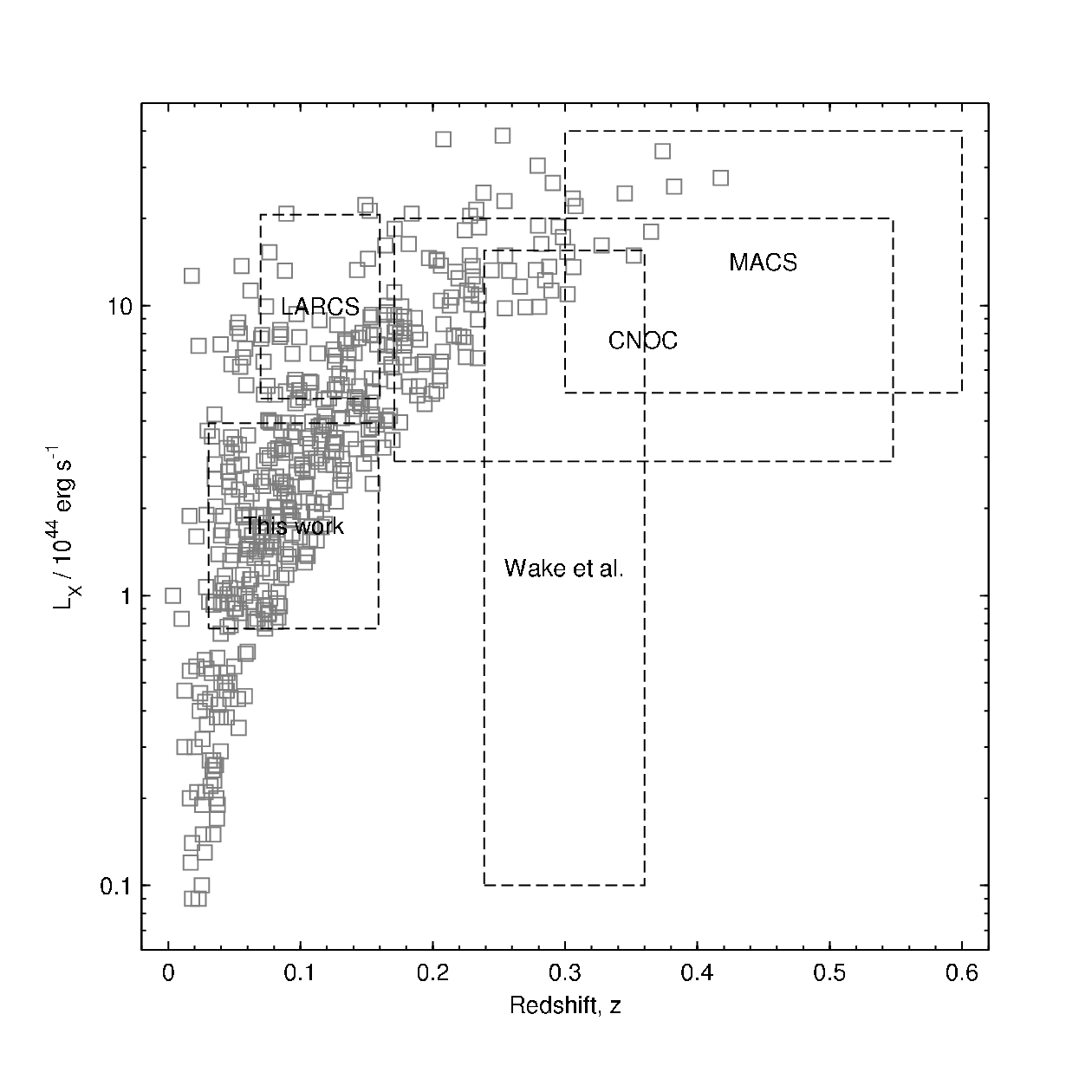,angle=0,width=3.5in}}
\vspace*{-0.7cm}
\caption{X-ray luminosity-redshift parameter space diagram. 
The grey squares represent XBACS, BCS, and eBCS clusters (Ebeling et al.\ 1996; 1998; 2000) 
while the dashed line boxes show the region of parameter space probed by a small variety X-ray 
selected cluster studies. Note that CNOC, 
MACS and Wake et al.\ select clusters from catalogues other than XBACS, BCS, and eBCS
-- only LARCS and this present study use the clusters contained inside the boxes displayed
and we emphasize that all works concerned attempt to minimize Malmquist bias.
The plot serves to 
demonstrate the bias to high $L_X$ clusters 
being visible to high redshift and here we
contend that we need a lower redshift, homogeneously selected 
baseline of intermediate-$L_X$ clusters to compare these other works to.
}
        \label{fig:lxz}
\end{figure}

Interpreting X-ray luminosity, $L_X$, as a proxy for cluster mass, 
the bias toward high mass is immediately apparent in a diagram of 
X-ray luminosity-redshift parameter space. In Fig.~\ref{fig:lxz}, 
we present the regions of this parameter space explored by four 
X-ray luminosity-selected galaxy cluster studies --
the Las Campanas/Anglo-Australian Telescope Rich Cluster Survey 
(LARCS; Pimbblet et al.\ 2001; 2002; 2006), 
the Canadian Network for Observational Cosmology (CNOC) 
Cluster Redshift Survey (Yee et al.\ 1996; Balogh et al.\ 1999), 
the Massive Cluster Survey (MACS; Ebeling et al.\ 2001) 
and the work of Wake et al.\ (2005) -- as well as the region explored by this work.

\begin{figure*}
\centerline{\psfig{file=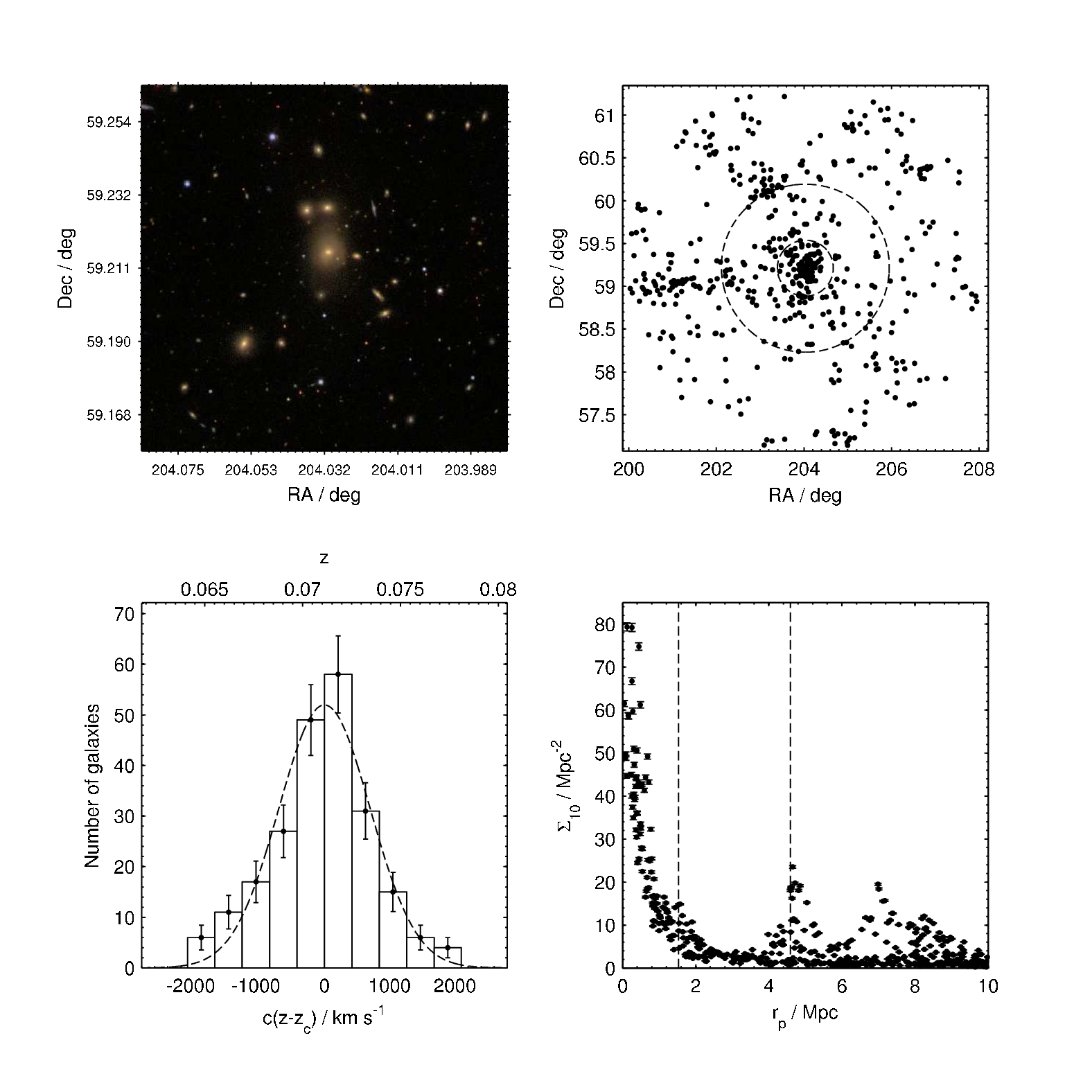,angle=0,width=6.75in}}
\vspace*{-0.7cm}
\caption{Diagnostic plots for Abell~1767 -- a representative cluster from our sample.
Top Left: SDSS image of the inner 0.5 $\times$ 0.5 Mpc of the cluster core.
Top Right: RA and Dec of all galaxies within 10 Mpc of the cluster centre.
The inner circle denotes $r_{200}$ and the outer circle is $3 r_{200}$.
Bottom Left: Redshift histogram for the cluster with the fitted Gaussian
overplotted that was used to determine the redshift and velocity dispersion
of the cluster.  Bottom Right: A diagnostic plot of local galaxy density ($\Sigma_{10}$)
as a function of projected radius from the cluster centre.  The two vertical
lines correspond to radii of $r_{200}$ and $3 r_{200}$.  Abell~1767 is a rich Bautz-Morgan
type II cluster fed by multiple filaments of galaxies (cf.\ Pimbblet et al.\ 2004)
with some substructure -- indicated by the peaks in $\Sigma_{10}$ that are prominent at $3 r_{200}$
and beyond.}
        \label{fig:a1767}
\end{figure*}

Given that luminosity-selected cluster surveys can be subject to such a 
Malmquist bias, a number of observational strategies are possible. The 
MACS and CNOC studies opt for a high $L_X$, longitudinal redshift strategy 
whereas the LARCS study focused on creating a homogeneous sample at low-redshift. 
In contrast, Wake et al.\ probes a wide range of X-ray luminosities at intermediate 
redshift by selecting clusters from a variety of surveys with different flux limits. 
A drawback of selecting clusters from multiple sources though, is that unforeseen 
biases may be introduced into the sample. Furthermore, such works do not tell 
us much about intermediate-mass clusters; for example, Wake et al.\ (2005) only 
looks at 12 clusters of which only 4 are found within the $L_X$ limits of 
this work -- comparable arguments can be extended to other studies (e.g.\ 
Huertas-Company et al.\ 2009). 
To address this issue, we assemble a sample of intermediate $L_X$ galaxy clusters 
with spectroscopically confirmed members from the Sloan Digital Sky Survey 
(SDSS Data Release 6, Adelman-McCarthy et al.\ 2008; see also York et al.\ 2000)
to examine the role of environment on the colour-magnitude relationship in
such systems.

On the colour-magnitude plane, there are two distinct galaxy populations 
residing in galaxy clusters. The first population is mainly composed of 
early-type (E and S0) galaxies which form a tight, linear relation 
extending to the brightest magnitude galaxies on this plane. 
This population is known as the red sequence and the ridge line 
upon which the red sequence galaxies lie is known as the colour-magnitude relation 
(CMR; Visvanathan \& Sandage 1977). Subsequent studies (e.g.\ Bower, Lucey \& Ellis 1992) 
have demonstrated the universal existence of a red sequence in all clusters. 
At faint magnitudes, the red sequence is observed to fan out on the 
colour-magnitude plane (Kodama \& Bower 2001; Pimbblet et al.\ 2002). 
This effect has been interpreted 
in terms of the age-metallicity relation --
the increasing spread of galaxy colours is the result 
of an increasing spread of galaxy metallicities and ages (Kodama et al.\ 1999; see also
Kodama \& Arimoto 1997). 
The second population is mainly composed of bluer, fainter-magnitude, morphologically
spiral and star-forming
galaxies which lie underneath the CMR -- the so-called blue cloud (see, e.g., 
Poggianti et al.\ 2006, Cortese \& Hughes 2009, Mei et al.\ 2009 and references therein). 
The evolution of this latter population in to the former has been 
extensively investigated in
the literature (e.g.\ Kodama \& Bower 2001; 
De Lucia et al.\ 2004; McIntosh et al.\ 2004;
Tanaka et al.\ 2005; Tran et al.\ 2005; Haines et al.\ 2006;
De Lucia et al.\ 2007; Stott et al.\ 2007; Ma et al.\ 2010;
Jim{\'e}nez et al.\ 2011) and relates to the origin of observed 
correlations such as the morphology-density relation (Dressler 1980)
and related issues (cf.\ Lewis et al.\ 2002; G{\'o}mez et al.\ 2003;
Butcher \& Oemler 1984; see also Mahajan \& Raychaudhury 2009).  
The developing picture is that as
low mass, blue galaxies accrete on to the cluster potential,
their star-formation is truncated 
(possibly after an intense starburst event; cf.\ Sato \& Martin 2006)
and their morphologies altered
(by a variety of physically-motivated mechanisms)
as they join more massive dark matter halos
(e.g.\ Gunn \& Gott 1972; 
Larson et al.\ 1980;
Byrd \& Valtonen 1990;
Moore et al.\ 1996;
Dressler et al.\ 1997; 
Quilis et al.\ 2000;
Balogh et al.\ 2000; 
Diaferio et al.\ 2001;
De Lucia et al.\ 2004).

In this context, a number of authors have presented evidence for
an environmental\footnote{The word `environment' 
has been taken to mean various things in the literature by different authors (see Haas, Schaye, 
\& Jeeson-Daniel 2011).
This includes but is not limited to: radius from a cluster centre, local galaxy density,
dark matter halo mass of a galaxy, and 
large-scale structural situation (e.g.\ void vs.\ filament vs.\ cluster).} 
dependence of galaxy observables.
For instance, Abraham et al.\ (1996) make use of a combined spectroscopic 
and photometric survey of Abell~2370 to show that radial gradients across the
cluster exist out to a remarkably large radius ($\sim 5$ Mpc).
In particular, the colours of CMR galaxies are shown to become progressively
bluer with projected radius (independent of mass)
from the cluster centre which is mirrored by
a change in Balmer line indices.
The natural interpretation of these results is that the mean 
luminosity-weighted age of the stellar populations in the cluster galaxies
is decreasing with increasing radius from the cluster centre as new galaxies
and groups are accreted from filaments, fuelling the cluster's growth 
(see also Bernardi et al.\ 2006; Smith et al.\ 2006).

Other studies followed Abraham et al.\ (1996) and found similar results
in other clusters
(e.g.\ Terlevich et al.\ 2001 for the Coma cluster; Haines et al.\ 2004).
Pimbblet et al.\ (2002; 2006) generalized these results for a very X-ray luminous
sample of $z\sim0.1$ galaxy clusters: a gradient of $d(B-R)/d r_p = -0.022 \pm 0.004$
was found for CMR galaxies out to 6 Mpc suggesting that galaxies in the outer regions
of the rich clusters have not yet had their star-formation rates fully truncated.
Concerned that such results may not be typical or representative of all clusters (since
Pimbblet et al.\ only examined the extrema of the cluster mass function),
Wake et al.\ (2005) extend this study to lower X-ray luminosities at modest redshifts
and found similar results (see their Fig.~13; see also Hogg et al.\ 2004).  
Yet, as noted above, this study only
constituted a small number of clusters at intermediate $L_X$.  Therefore in this work,
we aim to generalize these results to intermediate $L_X$ ranges through 
a purpose-constructed, representative sample of clusters.

The format of this work is as follows.
In Section~2, we construct a sample of 45 intermediate $L_X$ galaxy clusters at low redshift
that are covered by SDSS data.  We show
that these are representative of the general cluster population for these luminosities
and extract SDSS data for all member galaxies.  
In Section 3 we present Schechter function fits which we use to
define a characteristic magnitude for stacking galaxy clusters. We also present the
cluster colour-magnitude diagrams which are used to identify red sequence galaxies.
In Section~4, we stack the clusters to form a composite sample from
which we determine the photometric gradients of the red sequence galaxies in all SDSS
colours.  We describe how our results fit in with previous studies and suggest
the existence of relationships of modal colour gradient with $L_X$ and redshift.
Our major results are then summarized in Section~5.

Throughout this work, we adopt the standard flat cosmology from Spergel et al.\ (2007):
$\Omega_M = 0.24$, $\Omega_{\Lambda}=0.76$ and $H_0=73$ km s$^{-1}$ Mpc$^{-1}$.

\section{Data}

Galaxy clusters are selected for this work from the X-ray Brightest Abell Cluster Survey (XBACS), 
the ROSAT Brightest Cluster Sample (BCS) and the ROSAT Extended Brightest Cluster 
Survey (eBCS) catalogues (Ebeling et al., 1996; 1998; 2000). Briefly, the XBACS, 
BCS and eBCS catalogues list many extended, extragalactic X-ray sources from ROSAT 
All-Sky Survey data. Clusters are detected in the soft X-ray band between 
0.1--2.4 keV. 
In total, the catalogues contain 462 unique galaxy clusters of which 214 satisfy 
$0.7\times10^{44} < L_X < 4\times10^{44}$ ergs$^{-1}$ and $0.03<z<0.16$ -- our definition
of a local sample of intermediate $L_X$ galaxy clusters. 

\begin{table*}
\begin{center}
\caption{Intermediate $L_X$ galaxy cluster sample used in this work.}
\begin{tabular}{lccccccc}
\hline
Cluster & $L_X$ & B-M Type & Redshift & $\sigma_z$ & $r_{200}$ & C & S \\[-1ex]
& [10$^{44}$ erg/s] & & & [km/s] & [Mpc] & & \\[1ex]
\hline
Abell 602 & 1.14 & III        & 0.06195 $\pm$ 0.00006 & 1080 $\pm$ 40 & 2.39 $\pm$ 0.01 & 0.53  & yes \\
Abell 671 & 0.90 & II-III     & 0.04926 $\pm$ 0.00003 & 600 $\pm$ 10 & 1.34 $\pm$ 0.01 & 0.64 & ? \\
Abell 743 & 2.71 & III        & 0.13610 $\pm$ 0.00030 & 500 $\pm$ 200 & 0.93 $\pm$ 0.08 & 0.34  & no \\
Abell 744 & 0.77 & II         & 0.07298 $\pm$ 0.00009 & 240 $\pm$ 40 & 0.53 $\pm$ 0.02 & 0.65  & no \\
Abell 757 & 0.90 & III        & 0.05108 $\pm$ 0.00003 & 390 $\pm$ 10 & 0.88 $\pm$ 0.01 & 0.42 & no \\
Abell 763 & 2.34 & II-III     & 0.08924 $\pm$ 0.00008 & 380 $\pm$ 50 & 0.81 $\pm$ 0.02 & 0.43 & no \\
Abell 923 & 2.07 & II         & 0.11700 $\pm$ 0.00005 & 420 $\pm$ 30 & 0.88 $\pm$ 0.01 & 0.40 & no \\
Abell 957 & 0.78 & I-II       & 0.04546 $\pm$ 0.00005 & 850 $\pm$ 30 & 1.91 $\pm$ 0.01 & 0.55 & no \\
Abell 961 & 3.14 & II-III     & 0.12700 $\pm$ 0.00010 & 690 $\pm$ 40 & 1.43 $\pm$ 0.01 & 0.45 & no \\
Abell 971 & 1.44 & II         & 0.09320 $\pm$ 0.00002 & 762 $\pm$ 9 & 1.63 $\pm$ 0.01 & 0.78 & no \\
Abell 1035 & 0.92 & II-III    & 0.06721 $\pm$ 0.00005 & 720 $\pm$ 30 & 1.59 $\pm$ 0.01 & 0.61  & ? \\
Abell 1045 & 3.47 & II-III    & 0.13780 $\pm$ 0.00070 & 600 $\pm$ 100 & 1.28 $\pm$ 0.05 & 0.28 & no \\
Abell 1126 & 1.15 & I-II      & 0.08433 $\pm$ 0.00007 & 350 $\pm$ 50 & 0.75 $\pm$ 0.02 & 0.46 & no \\
Abell 1361 & 3.59 & I-II      & 0.11549 $\pm$ 0.00005 & 640 $\pm$ 20 & 1.33 $\pm$ 0.01 & 0.45 & no \\
Abell 1446 & 1.30 & II-III    & 0.10280 $\pm$ 0.00010 & 710 $\pm$ 60 & 1.51 $\pm$ 0.02 & 0.52  & no \\
Abell 1691 & 0.89 & II        & 0.07176 $\pm$ 0.00002 & 650 $\pm$ 8 & 1.42 $\pm$ 0.01 & 0.52  & yes \\
Abell 1728 & 1.29 & I-II*     & 0.08965 $\pm$ 0.00002 & 940 $\pm$ 10 & 2.02 $\pm$ 0.01 & 0.34 & yes \\
Abell 1767 & 2.47 & II        & 0.07112 $\pm$ 0.00002 & 700 $\pm$ 20 & 1.53 $\pm$ 0.01 & 0.54  & ? \\
Abell 1773 & 1.53 & III       & 0.07727 $\pm$ 0.00003 & 790 $\pm$ 40 & 1.72 $\pm$ 0.01 & 0.53 & yes \\
Abell 1809 & 1.61 & II        & 0.07938 $\pm$ 0.00001 & 808 $\pm$ 7 & 1.76 $\pm$ 0.01 & 0.76  & ? \\
Abell 1814 & 2.82 & II        & 0.12673 $\pm$ 0.00009 & 330 $\pm$ 50 & 0.68 $\pm$ 0.02 & 0.72  & no \\
Abell 1831 & 1.90 & III       & 0.06311 $\pm$ 0.00003 & 480 $\pm$ 10 & 1.06 $\pm$ 0.01 & 0.52  & yes \\
Abell 1885 & 2.40 & II-III    & 0.09200 $\pm$ 0.00200 & 1100 $\pm$ 500 & 2.40 $\pm$ 0.20 & 0.80  & ? \\
Abell 1925 & 1.56 & II        & 0.10570 $\pm$ 0.00007 & 580 $\pm$ 30 & 1.22 $\pm$ 0.01 & 0.34 & ? \\
Abell 1927 & 2.14 & I-II      & 0.09506 $\pm$ 0.00003 & 520 $\pm$ 20 & 1.12 $\pm$ 0.01 & 0.37  & no \\
Abell 2033 & 2.57 & III       & 0.07828 $\pm$ 0.00003 & 1049 $\pm$ 9 & 2.28 $\pm$ 0.01 & 0.51 & yes \\
Abell 2108 & 1.97 & III       & 0.09033 $\pm$ 0.00008 & 750 $\pm$ 30 & 1.60 $\pm$ 0.01 & 0.51  & no \\
Abell 2110 & 3.93 & I-II      & 0.09728 $\pm$ 0.00005 & 630 $\pm$ 40 & 1.35 $\pm$ 0.02 & 0.42 & no \\
Abell 2124 & 1.35 & I         & 0.06723 $\pm$ 0.00002 & 740 $\pm$ 10 & 1.62 $\pm$ 0.01 & 0.68  & no \\
Abell 2141 & 3.89 & II        & 0.15900 $\pm$ 0.00030 & 900 $\pm$ 300 & 1.80 $\pm$ 0.10 & 0.80 & no \\
Abell 2148 & 1.39 & III*      & 0.08843 $\pm$ 0.00005 & 570 $\pm$ 20 & 1.22 $\pm$ 0.01 & 0.57  & ? \\
Abell 2149 & 0.83 & II-III*   & 0.06503 $\pm$ 0.00003 & 280 $\pm$ 20 & 0.61 $\pm$ 0.01 & 0.35  & no \\
Abell 2175 & 2.93 & II        & 0.09646 $\pm$ 0.00004 & 780 $\pm$ 10 & 1.65 $\pm$ 0.01 & 0.44 & no \\
Abell 2199 & 3.70 & I         & 0.03055 $\pm$ 0.00001 & 681 $\pm$ 1 & 1.56 $\pm$ 0.01 & 0.43  & yes \\
Abell 2228 & 2.81 & I-II      & 0.10102 $\pm$ 0.00004 & 980 $\pm$ 20 & 2.09 $\pm$ 0.01 & 0.41  & no \\
RXJ0820.9+0751 & 2.09 & I-II* & 0.11032 $\pm$ 0.00007 & 560 $\pm$ 30 & 1.17 $\pm$ 0.01 & 0.54  & no \\
RXJ1000.5+4409 & 3.08 & III*  & 0.15310 $\pm$ 0.00030 & 700 $\pm$ 100 & 1.33 $\pm$ 0.05 & 0.82  & ? \\
RXJ1053.7+5450 & 1.04 & III*  & 0.07291 $\pm$ 0.00004 & 580 $\pm$ 20 & 1.27 $\pm$ 0.01 & 0.62  & yes \\
RXJ1423.9+4015 & 0.94 & III*  & 0.08188 $\pm$ 0.00005 & 460 $\pm$ 10 & 0.99 $\pm$ 0.01 & 0.54  & ? \\
RXJ1442.2+2218 & 2.66 & II*   & 0.09613 $\pm$ 0.00007 & 470 $\pm$ 70 & 1.01 $\pm$ 0.03 & 0.42  & ? \\
RXJ1652.6+4011 & 2.85 & III*  & 0.14800 $\pm$ 0.00020 & 750 $\pm$ 90 & 1.52 $\pm$ 0.03 & 0.45  & yes \\
ZwCl 1478 & 2.38 & II-III*    & 0.10300 $\pm$ 0.00080 & 400 $\pm$ 200 & 0.82 $\pm$ 0.08 & 0.35  & no \\
ZwCl 4905 & 1.20 & I-II*      & 0.07641 $\pm$ 0.00003 & 480 $\pm$ 20 & 1.04 $\pm$ 0.01 & 0.48  & no \\
ZwCl 6718 & 1.24 & II*        & 0.07220 $\pm$ 0.00020 & 400 $\pm$ 200 & 0.85 $\pm$ 0.09 & 0.55 & no \\
ZwCl 8197 & 2.89 & II*        & 0.11320 $\pm$ 0.00010 & 830 $\pm$ 100 & 1.74 $\pm$ 0.04 & 0.31 & no \\
\hline
\noalign{\smallskip}
\end{tabular}
  \label{tab:clusters}
\end{center}
\end{table*}

\begin{figure}
\centerline{\psfig{file=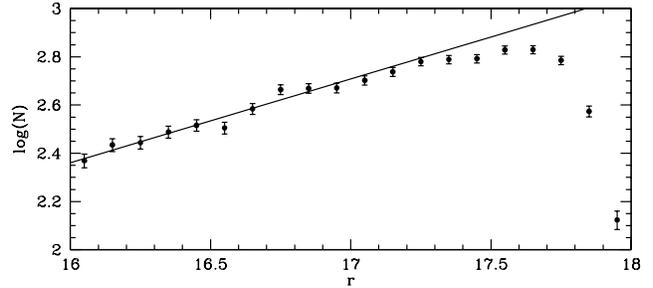,angle=0,width=3.5in}}
\vspace*{-5cm}
\caption{Histogram of r-band magnitudes for our galaxy sample. To determine
how complete our sample is, we fit a line to the brighter part of this plot
($16<r<17$) where we are confident of being 100 per cent complete and extrapolate
this to fainter magnitudes.  At $r=17.77$, our sample is still 95 per cent
complete and we suggest that this level of completeness is sufficient for
the present work.
}
        \label{fig:rhistlog}
\end{figure}

We use SDSS DR6 (Adelman-McCarthy et al.\ 2008) as our photometric and spectroscopic
catalogue in this work, which immediately cuts down the number of potential clusters
available to study due to the spatial coverage of SDSS and also restricts our
coverage to modestly bright magnitudes. The SDSS 
main sample spectroscopic target list contains all galaxies
brighter than $r=17.77$ (Strauss et al.\ 2002) but due to fibre collisions
(cf.\ Blanton et al.\ 2003), the completeness level at $r=17.77$ may be
slightly lower than 100 per cent at this magnitude.  In Fig.~\ref{fig:rhistlog}
we display our own estimate of the completeness of our catalogue and find
that at $r=17.77$, we are still $\sim 95$ per cent complete which should be more
than sufficient for the present work. 

Preliminary cluster membership is determined by extracting galaxies 
within $z\pm0.02$ of the nominal cluster redshift given by XBACS, BCS and eBCS out to a 
projected radius of 10 Mpc from the cluster centre (defined as the X-ray luminosity
peak of the cluster).  
For each cluster, we determine final membership by fitting a Gaussian
to the redshift distribution of the preliminary members (Fig.~\ref{fig:a1767}). 
Using the position of the peak of the Gaussian as the updated cluster redshift,
we proceed to eliminate any galaxies outside $3 \sigma_z / c$ of the updated
redshift (i.e.\ a sigma-clipping process;
cf.\ Yahil \& Vidal 1977).  This process is iterated twice to produce the 
final membership for the clusters in our sample.  Although we may miss
real cluster members through this method (cf.\ section 4.2 of Pimbblet et al.\ 2006),
it will suffice for our purposes as we will later stack all of our clusters
together to form a composite.

To ensure that this method is robust against choice of bin size in 
constructing the redshift histogram,
Gaussian fit coefficients for the redshift histograms are estimated 
iteratively by performing $\chi^2$
minimizations on the histogram data points. One hundred iterations 
are performed for each cluster with new random bin sizes on each 
iteration. The best fit Gaussian coefficients and their associated 
errors are calculated from the mean and standard deviation values 
of the coefficients over 100 iterations. By randomising the bin sizes, 
we ensure that our best fit Gaussians are independent of binning bias.

Our examination of the clusters generated by this method (e.g., Fig.~\ref{fig:a1767})
returns a mixture of usable and unusable clusters. 
Here, we consider `unusable' to be clusters with fewer than 50 members (this is driven
by a need to maintain a high completeness in each cluster so that 
we can fit a line to the colour-magnitude relation of an individual cluster and
so that an individually rich cluster will not dominate in a stacked, composite sample),
those located in the `whisker' regions of SDSS DR6 (i.e.\ those
contained in the long, narrow stripes of spectroscopically-sampled sky near the celestial equator
where we cannot probe out to $3r_{200}$) and over-lapping clusters.  
For the former issue, this introduces a minor bias to our target selection: low redshift
clusters are much more able to enter our sample than higher redshift clusters 
(clusters at $z<0.1$ have $182 \pm 48$ extra members compared to 
$z>0.1$) meaning that some of the advantage our sample
has (Fig.~1) is removed.  However, if we were to add in clusters 
with $<50$ members,
their contribution to the final, composite cluster would be 
fractional (one extra cluster with $<50$ members would increase our
sample size by $<<0.5$ per cent) and our final results would not 
be significantly affected.

We define overlapping clusters as clusters which have their 10 Mpc projected
radius search areas overlap with the search areas of other clusters within our
$L_X$ range.  Removing 
the overlapping clusters simplifies our analysis because it means that we do 
not have to worry about how to assign cluster membership to galaxies located 
in the overlap regions and cross-contamination
(explicitly, an overlapping neighbour 
will manifest itself as a second red sequence in a colour-magnitude diagram which will
mean any fit to the colour-magnitude relation of an individual cluster is suspect at best).
We caution that there may still be clusters overlapping with other clusters
that are outside of our $L_X$ range but we would anticipate that their influence on
our result to be minimal when we later combine our clusters in to a composite sample.

Our final sample consists of 45 intermediate $L_X$ galaxy clusters 
at a median redshift of $z=0.0897$ which span a range of properties.
For example, our richest cluster (Abell 2199) 
contains 1344 spectroscopically-confirmed members whereas our poorest 
cluster (Abell 1814) has 54 members.
We summarize our cluster sample and their global
properties in Table~\ref{tab:clusters}.

To increase the utility of our analysis to 
other researchers (and our own future work), we derive several additional
parameters for the clusters used in this work.
Firstly, included in Table~\ref{tab:clusters} are Bautz-Morgan classifications 
of our clusters (Bautz \& Morgan 1970).  Most of these are directly 
sourced from the NASA Extragalactic Database (NED), but those marked
by an asterisk were manually determined by visual inspection by PCJ.
Values for $r_{200}$ are derived following Carlberg et al.\ (1997):
$R_{Virial} \approx r_{200} = \sqrt(3) \sigma_z / 10 H(z)$ 
where 
$H(z) = H_0^2 (1+z)^2 (1+\Omega_M z)^2$ 
and
$\sigma_z$ is the velocity dispersion of the cluster.
The concentration indices ($C$) are computed by sorting 
all galaxies within $3r_{200} \approx r_{100}$ in to 
projected radius order and applying 
$C = log_{10} (r_{60} / r_{20})$ where $r_{60}$ and $r_{20}$
are the 20th and 60th percentiles of this distribution.
We also make a qualitative judgment about whether the cluster
has substructure (column headed $S$ in Table~\ref{tab:clusters}) from inspection of a plot of local galaxy
density\footnote{We define local galaxy density as $\Sigma_{10}$ -- 
the surface area on the sky that is occupied by a given galaxy and its 10 nearest neighbours.
This measure of `environment' probes the internal densities of the dark matter halos
well (Muldrew et al.\ 2011).}
versus projected radius (Fig.~\ref{fig:a1767}).  A peak outside of the cluster core in $\Sigma_{10}$
is taken as a signature of substructure for this purpose.
Although in principle we could have used other, quantitative methods (cf.\ Dressler \& Shectman 1988;
Ashman et al.\ 1994; see also Pimbblet et al.\ 2011 and references therein)
this approach is sufficient to make the statement that our sample has a range of clusters at
a number of evolutionary stages.
Including both clusters that have strong secondary $\Sigma_{10}$ peaks and those with
probable ones (i.e.\ both `yes' and `?' entries in the substructure column, $S$, 
of Table~\ref{tab:clusters}), 
our sample has 19 out of 45 clusters ($\sim 40$ per cent) 
containing substructure.  This is an under-estimate for intermediate $L_X$ clusters
overall since we have intentionally removed overlapping clusters to generate 
our sample (above) and we don't search for line-of-sight substructure.
We note that this value is intermediate to that predicted for rich clusters
(e.g., Lacey \& Cole 1993) and that observed in poor clusters (Burgett et al.\ 2004).
This is in line with theoretical expectations that demonstrate   
intermediate-$L_X$ clusters are expected to have accreted approximately 50 per cent 
of their total mass in the past $\sim 7$ Gyr (see Fig.~13 of Lacey \& Cole 1993).

\begin{figure*}
\centerline{\psfig{file=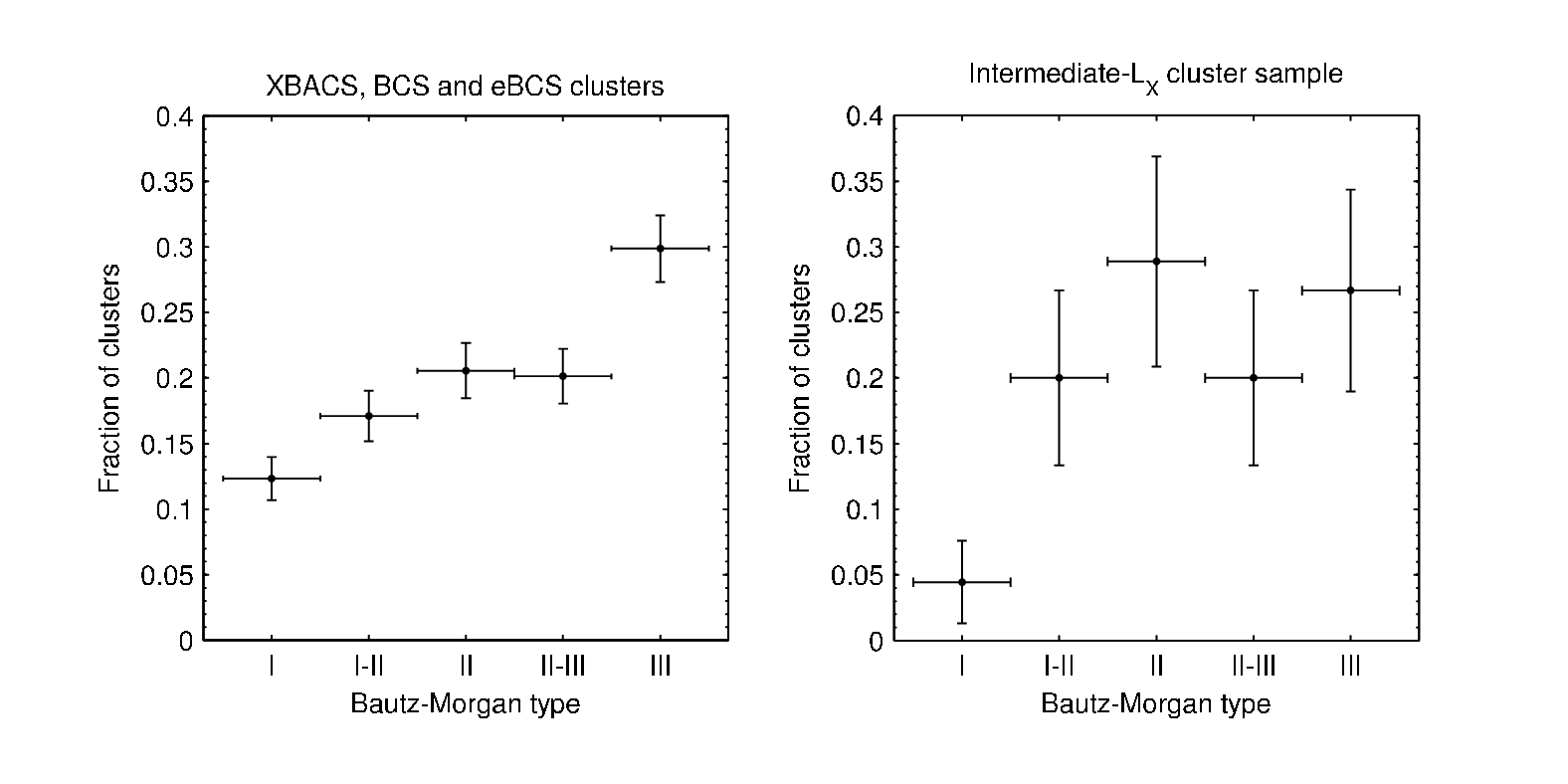,angle=0,width=6.75in}}
\vspace*{-0.7cm}
\caption{Bautz-Morgan fractions for the XBACS, BCS, and eBCS
catalogues (462 clusters) and our cut-down intermediate $L_X$ cluster sample (45 clusters).
The vertical error bars are Poissionian.  The two samples are not significant different at a $3\sigma$ level.
}
        \label{fig:BMfracs}
\end{figure*}

To characterize the dataset and address the question of whether our sample of 45 galaxy clusters
is representative of the global intermediate $L_X$ cluster population we compare 
the frequency of different
Bautz-Morgan classifications in our sample and the parent XBACS, BCS
and eBCS sample.
This is graphically displayed in Fig.~\ref{fig:BMfracs}. In comparison to
the full X-ray sample of intermediate $L_X$ clusters, our cut-down sample appears 
to have a lower prevalence of BM I and slightly higher 
prevalence of BM II types. 
The fraction of BM III (and, indeed, intermediate BM types) appears 
to be similar in both populations. Despite this, the differences 
between the histograms are not significant at the $3\sigma$ level.
We therefore contend that we have a representative sample of all intermediate $L_X$ cluster
types for this work.
However, from Fig.~\ref{fig:lxz} it is also apparent that we are not probing the lower
right of the box denoting this work (i.e.\ clusters at the upper end of our redshift
interval with low $L_X$). In acknowledging that this biases our cluster sample, we note the 
lookback time from the lowest redshift to the highest is some $\sim 1.5$ Gyr.

\section{Luminosity Function and Colour-Magnitude Relations}
\begin{figure*}
\centerline{\psfig{file=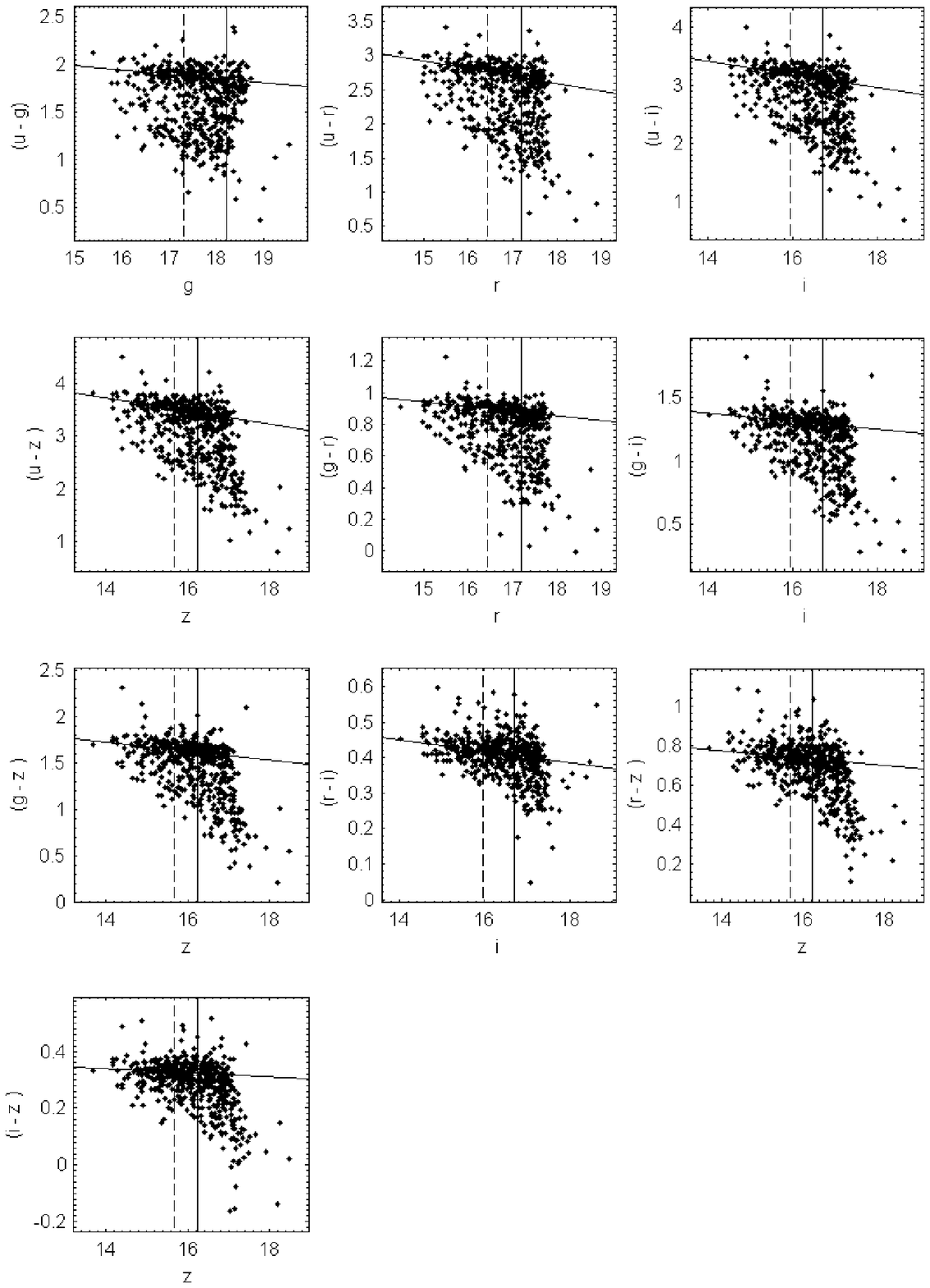,angle=0,width=6.5in}}
\vspace*{-1.5cm}
\caption{Colour-magnitude diagrams for Abell~1767 in a variety of broadbands
for all member galaxies within 10 Mpc of the cluster centre (see Fig.~\ref{fig:a1767}).
For each diagram, we fit the red sequence (diagonal solid line) as described in the text.
The vertical lines denote M$^{\star}$ (dashed line) and our fiducial magnitude
(solid line) which represents a 90 per cent completeness limit for our sample (see
Fig.~\ref{fig:rhistlog}).
We use $M^{\star}$ as the point about which we pivot the plots to create the composite cluster.
Analogous diagrams are constructed for the other clusters in our sample.
}
        \label{fig:cmrs}
\end{figure*}

Having defined our cluster sample and their galaxy members, we now create
plots of colour-magnitude relationships for all of our clusters in
a variety of broadbands (Fig.~\ref{fig:cmrs}).
In creating these colour-magnitude diagrams, we located a number of 
discordant points -- galaxies that lie very far away from the bulk of the 
population.  These galaxies are explored in detail in Appendix~A but here
we simply note that they are removed from the subsequent analysis as they
may adversely affect the fit of the CMRs.

We first fit the absolute (k-corrected) magnitudes of the
galaxy cluster members with a Schechter function (Schechter 1976)
to determine $M^{\star}$ (a more physically meaningful parameter than
absolute magnitudes)
and therefore provide a pivot point which we will
use to combine our CMRs together in to a stacked cluster (dashed vertical
lines in Fig.~\ref{fig:cmrs}).  
The k-corrections used in this work are from 
the {\sc photoz} table of SDSS DR6 
(http://cas.sdss.org/astro/en/help/browser/browser.asp?
n=Photoz\%20\&t=U).
Although these use photometric redshifts, the redshift contribution to the total
error is insignificant in comparison to the error budget of the flux density
which is rarely known to better than a few per cent (Hogg et al.\ 2002).
For each absolute magnitude band, the Schechter functions are fit from 
the brightest magnitude bin up to the 90 per cent completeness limit (solid vertical
line in Fig.~\ref{fig:cmrs}; see Fig.~\ref{fig:rhistlog} for the definition).
One hundred $\chi^2$ minimizations are performed with $M^{\star}$, $N^{\star}$ 
and $\alpha$ as free parameters and with randomised bin sizes between 0.05 and 
0.5 mag. The fit parameters for $M^{\star}$ 
and their associated errors were calculated as being 
the mean and standard deviation of the 100 runs. These values for
$M^{\star}$ (critical to this study) are shown in Table~\ref{tab:lf}.
We note that single Schechter function fits are inadequate 
descriptions of the luminosity function at both the bright end (where double 
or Gaussian functions may be more appropriate; e.g.\ 
Thompson \& Gregory 1993; Dahl{\'e}n et al.\ 2004) and the faint end (since 
we truncate our fits to the 90 per cent completeness limit).  They do,
however, provide an excellent fit to the `knee' of the distribution
and consequentially $M^{\star}$ which is important to us as a pivot point since
it does not significantly vary with redshift (Mobasher et al.\ 2003 and references
therein).  The values of $M^{\star}$ that we present
in Table~\ref{tab:lf} are similar to other values
reported for generic SDSS cluster populations (Popesso et al.\ 2005), but we
refrain from a more detailed comparison of these luminosity functions since
our cluster sample is constructed in a very different manner to those
studies and with different goals in mind.

\begin{table}
\begin{center}
\caption{Values for $M^{\star}$ from fitting Schechter functions to the
cluster galaxies.
\hfil}
\begin{tabular}{ll}
\noalign{\medskip}
\hline
Band & $M^{\star}$ \\ 
\hline
$u$ & -18.29$\pm$0.04 \\ 
$g$ & -20.35$\pm$0.05 \\ 
$r$ & -21.09$\pm$0.05 \\ 
$i$ & -21.53$\pm$0.04 \\ 
$z$ & -21.81$\pm$0.04 \\ 
\hline
\end{tabular}
  \label{tab:lf}
\end{center}
\end{table}

To fit the CMRs, we only use galaxies brighter than the 90 per cent completeness
fiducial magnitudes (vertical solid line in Fig.~\ref{fig:cmrs}).
Since there are two overlapping galaxy populations on the colour-magnitude 
plane (red sequence and blue cloud), ordinary least squares fitting techniques 
are inadequate to fit the CMRs. Typically, ordinary least squares fitting yields 
slopes which are too steep and intercepts which are too high -- 
the faint end of the CMR being dragged blueward by the blue cloud. 
Ideally, one would like to remove the blue cloud galaxies and perform 
an ordinary least squares fit to the remaining red sequence galaxies. 
However, identifying blue cloud galaxies without already knowing the 
position of the CMR is a non-trivial exercise. A more straightforward 
approach is to perform a robust linear fit which effectively weights 
red sequence galaxies higher than blue cloud galaxies (see for example Pimbblet et al.\ 2002
who use a merit function which minimizes the sum of the absolute values of the residuals).
Here, we follow a suggestion by Press et al.\ (1992) and use the Lorentzian merit 
function:
\begin{equation}
\sum_i \frac{log(1+r^2_i / 2)}{\Delta y_i}
\end{equation}
where $\Delta y_i$ is magnitude of the error in y. 
In comparison to the absolute deviation minimizing 
merit function, the Lorentzian merit function visually 
appears to converge at least as well or closer to the 
apparent red sequence. The merit function was minimized 
using the Nelder-Mead downhill simplex algorithm (Press et al.\ 
1992). 
The Nelder-Mead algorithm requires initial guess values for the 
linear fit coefficients -- we use ordinary least squares regression fit 
coefficients as our starting points.

In almost all cases our approach 
successfully located the CMR in each diagram.
However, the algorithm failed for a few clusters.  In these cases, we
impose a colour envelope around the (eyeball determined) location of the CMR
to help the algorithm hone in on the CMR on a second run.
All fits are double-checked by eye to confirm the approach works (Fig.~\ref{fig:cmrs}).

\section{Composite cluster and Discussion}
We now exploit the homogeneity of our sample and stack our clusters together
to form a composite.
This is achieved by applying colour and magnitude 
transformations to the CMRs of each of our clusters (Fig.~\ref{fig:cmrs})
in a two-step process. 
Firstly, we remove the slope of the CMR and 
secondly, we evolve the individual clusters to a common redshift.

For each cluster we identify a pivot point which we define to be the 
intersection of the CMR with $M^{\star}$.
We rotate every point on the colour-magnitude diagram about this
pivot point such that the CMR becomes horizontal.  We emphasize that 
the slopes of the CMRs of the individual clusters have been 
found in an homogenous manner for our sample -- an absolute requirement
given the uncertainties (cf.\ Pimbblet et al.\ 2002).

\begin{figure}
\centerline{\psfig{file=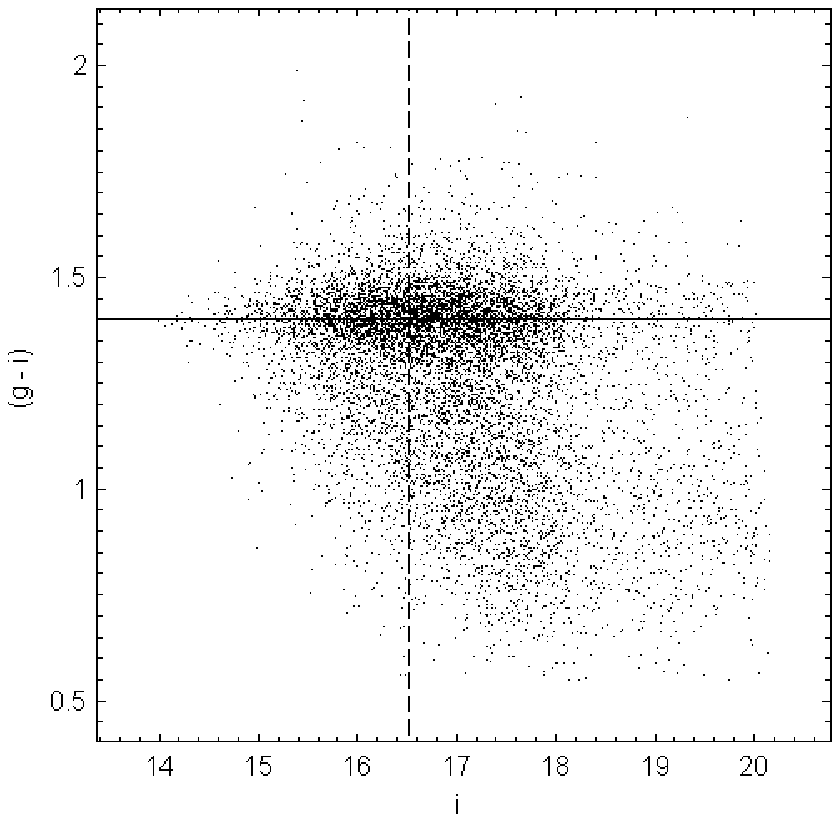,angle=0,width=3.75in}}
\vspace*{-0.7cm}
\caption{Composite $z\sim0.09$ cluster colour-magnitude diagram in (g-i) versus i
for all 45 clusters in our sample.  
The solid horizontal line denotes the transformed position of the CMR
of all the individual clusters whilst the vertical dashed line is $M^{\star}$.
}
        \label{fig:composite}
\end{figure}

We then evolve the
clusters to our mean redshift ($z = 0.0897$) by performing a vertical
colour shift and an horizontal magnitude shift. The colour scale of
each cluster is shifted so that the red sequence ridge lines are
stacked on top of each other at a colour consistent with $M^{\star}$ 
at the mean redshift of the sample. 
The magnitude scale of each cluster is also
shifted so that the $M^{\star}$ value of each cluster coincides.
Fig.~\ref{fig:composite} displays the colour-magnitude diagram of the composite cluster in
(g-i) versus i.  Analogous plots are created for the other colour versus
magnitude combinations permitted by the SDSS data.

We now turn to the environmental dependence of the modal colour of the CMR.
We define modal colour to be the colour corresponding to the peak of
a Gaussian fitted to the CMR peak of colour histogram of the composite cluster.
To probe modal 
colour dependence as a function of cluster environment, we employ a 
method similar to the one described in Pimbblet et al.\ (2002; 2006).
Briefly, this is done by dividing the composite CMR (Fig.~\ref{fig:composite})
into projected radius (via a fixed metric) and local galaxy density bins
down to a fixed absolute magnitude limit ($M^{\star}+1$).
Although we could have chosen $r_{200}$ to place our clusters on to a more
physically-motivated spatial scale, we choose to use a fixed metric since
$r_{200} \approx r_{Virial}$ has only a weak dependence on X-ray luminosity
(Babul et al.\ 2002; see also Pimbblet et al.\ 2002).  Given the narrow
$L_X$ range of our sample, the variation of $r_{200}$ for the cluster sample
is predicted to be small 
(indeed, this prediction is bourne out in our calculations presented in 
Table~\ref{tab:clusters}) hence the clusters
are scaled to the fixed metric for simplicity.

\begin{figure*}
\centerline{
\psfig{file=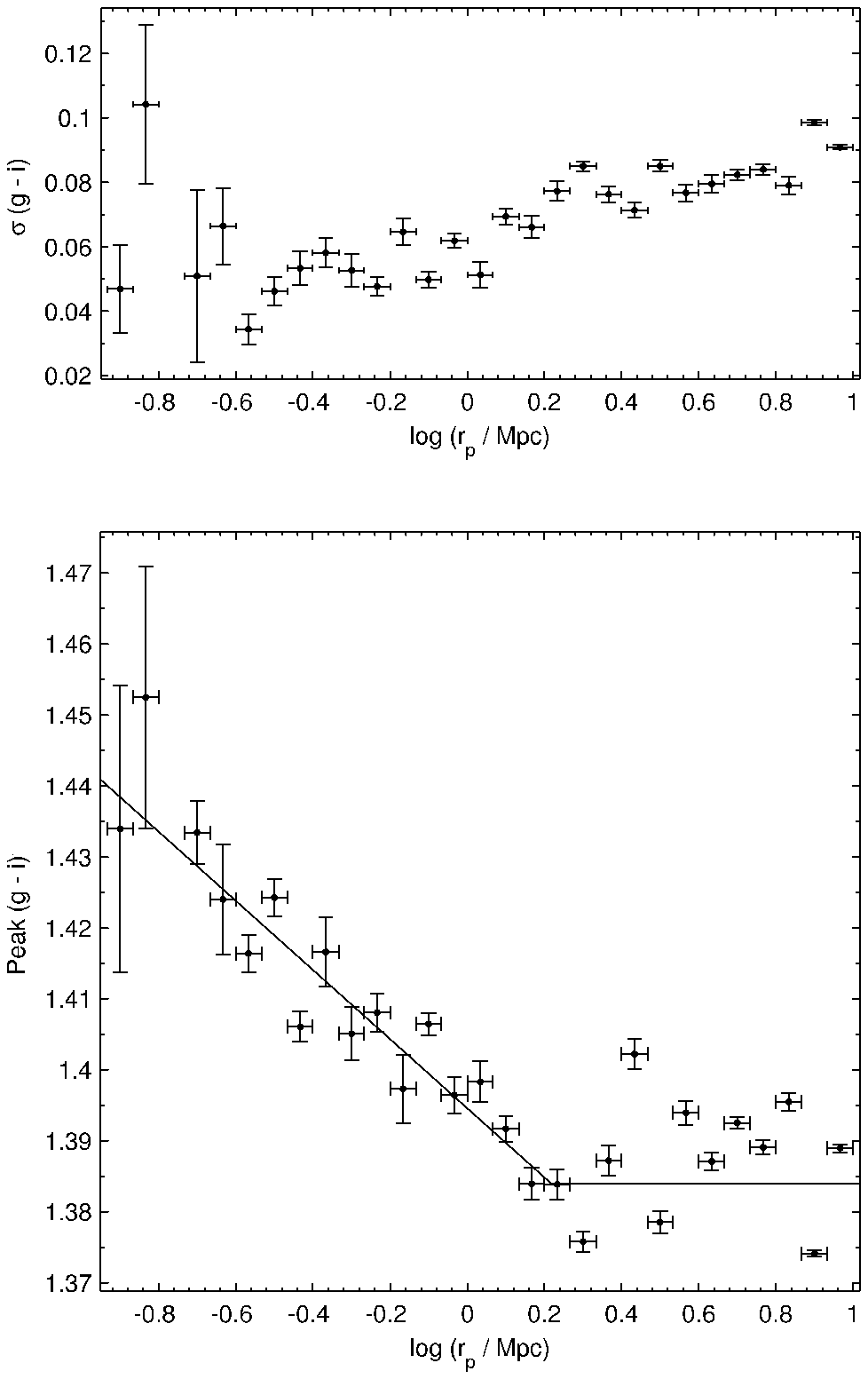,angle=0,width=4.5in}
\hspace*{-2.5cm}
\psfig{file=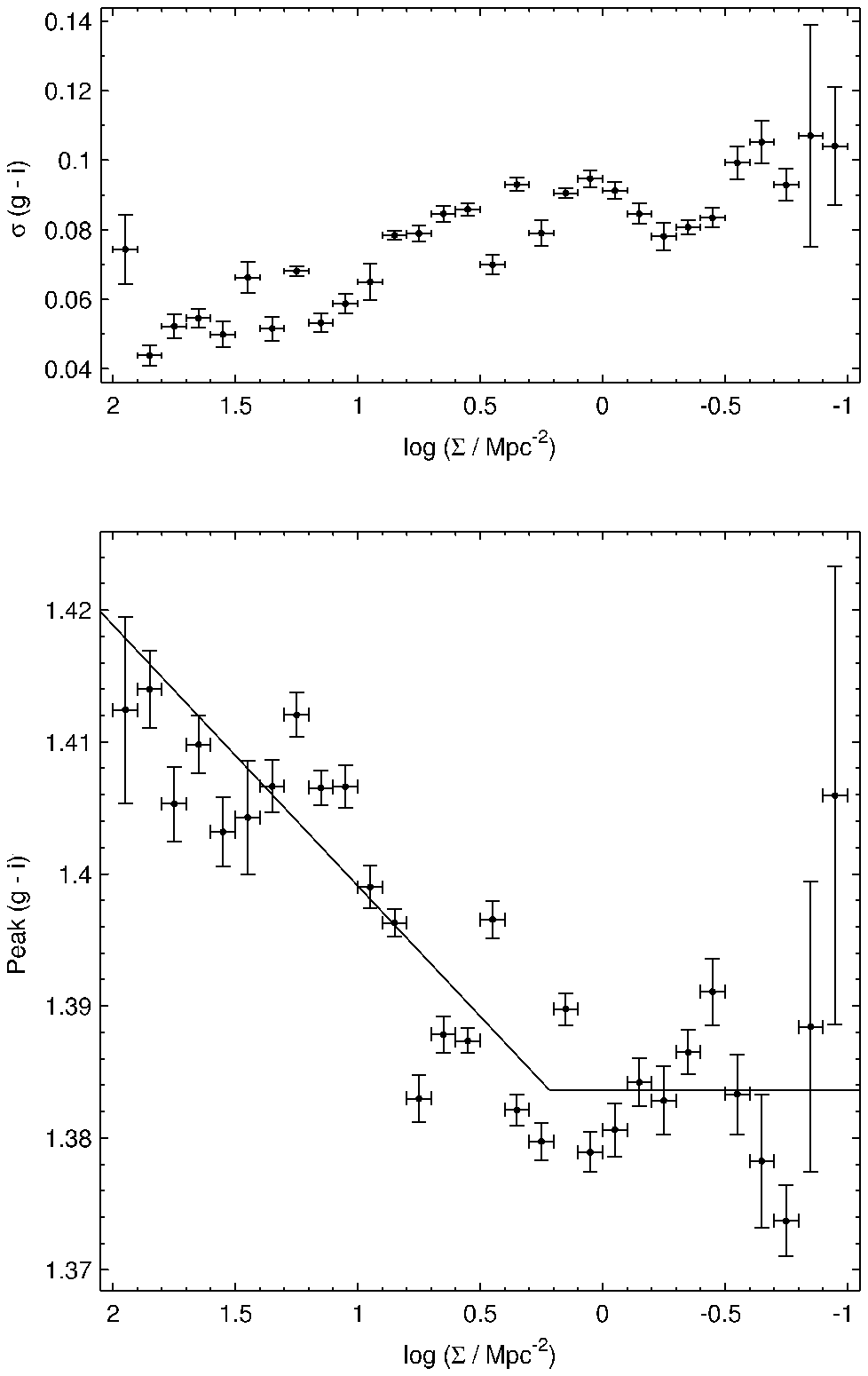,angle=0,width=4.5in}
}
\vspace*{-2.5cm}
\caption{Environmental dependence of the width (upper panels)
and modal value of the red sequence peak (lower panels).
The colours of the CMR galaxies become progressively bluer with increasing radius
from the cluster centre and decreasing local galaxy density (the lines
in the lower panels are two linear fits to the points, divided where the gradient becomes
approximately zero).  
This is accompanied with an increase in the width of the CMR.
}
        \label{fig:depends}
\end{figure*}

In Fig.~\ref{fig:depends} we present the dependence of the modal CMR value 
on these two parameters.  Qualitatively in-line with previous works (Abraham et al.\ 1996;
Terlevich et al.\ 2001; Pimbblet et al.\ 2002; Wake et al.\ 2005; 
Pimbblet et al.\ 2006; see also Girardi et al.\ 2003; Haines et al.\ 2004),
we observe the modal red sequence 
colour become progressively bluer with radius away from the
cluster centre and, equivalently, decreasing local galaxy density.
At the same time, the width of the red sequence is observed to increase.
Similar results are obtained for the other SDSS bands.
The strongest gradient in modal colour occurs for the inner portions (equivalently,
high density regions) of these plots (see Fig.~\ref{fig:depends}).
We interpret the increasing width and bluening of the CMR 
as an age-metallicity effect (Kodama \& Arimoto 1997; 
Kodama \& Bower 2001; Pimbblet et al.\ 2006) whereby
the red sequence galaxies on the outskirts of the
clusters have younger luminosity-weighted ages than those 
at the core.
One potential issue with the increasing width is that as one
moves from the cluster core to its outskirts, the mixture
of the galaxies may be changing.  For instance, the typical galaxy 
at higher radius may have a lower mass. This would result in larger
error bars on the colour of the galaxy and could naturally explain 
the increase in width as simply being a relic of measurement errors.
The colour error budget is, however, dominated by the fainter
galaxies at all radii (simply due to their numbers).
The median (g-r) colour error in our composite cluster 
at $<1$ Mpc is 0.009, which increases to 0.010 at 5--6 Mpc.
These figures $\sim$halve if we consider only galaxies with
$M_r<-21$ and increase by some $\approx$0.005 for $M_r>-20$
galaxies.  In all cases, the change in the photometric error
with radius from the cluster centre is much less than the CMR
width (Fig.~\ref{fig:depends}) and at most contributes a quarter 
of the CMR's width (i.e.\ at low radii).
We also check the reality of the radial bluening by isolating
samples in different absolute magnitude ranges (i.e.\ a mass proxy).
In all cases, a blueward trend of modal CMR colours is found.

\begin{figure*}
\centerline{\psfig{file=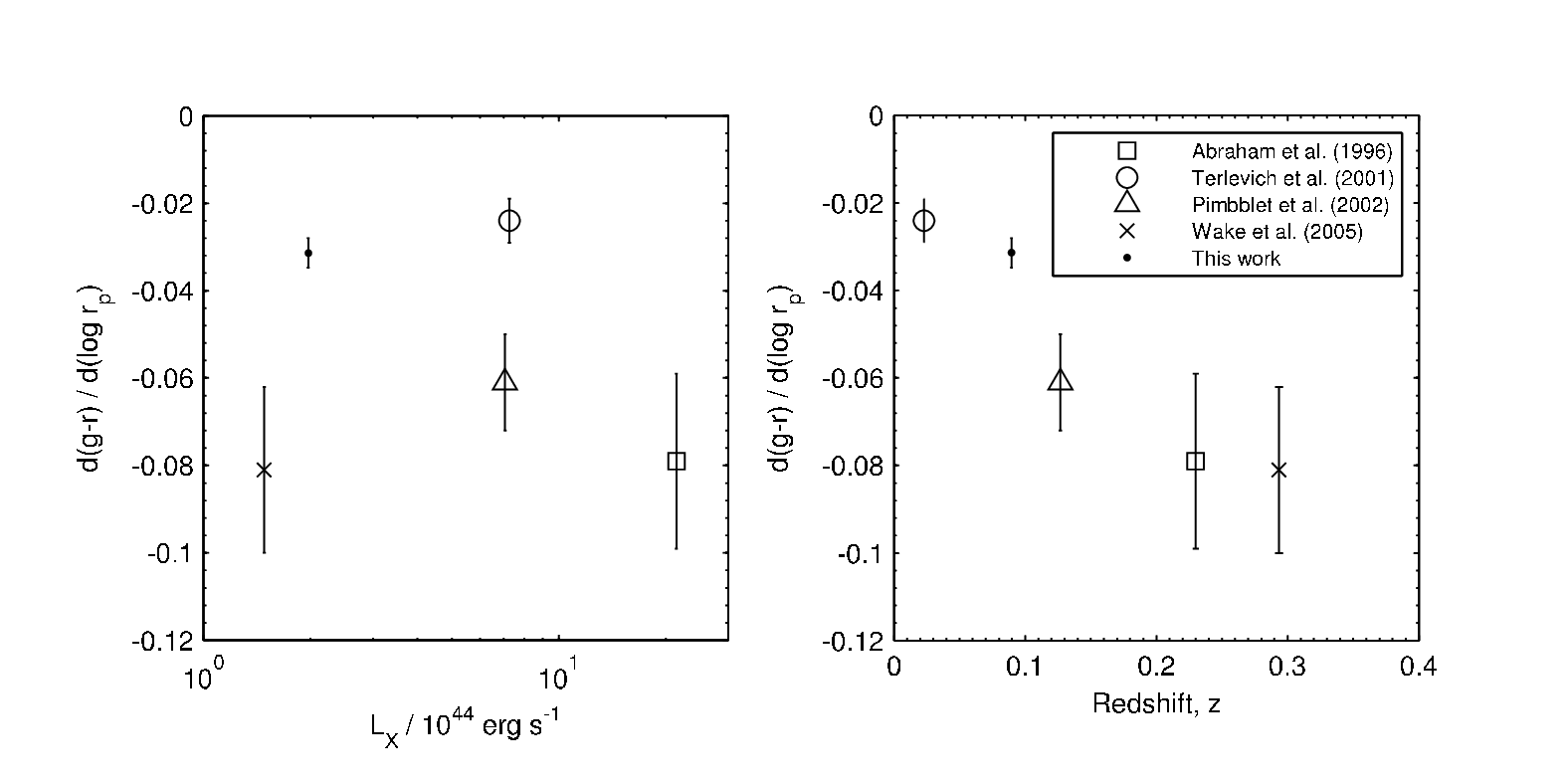,angle=0,width=6.75in}}
\vspace*{-0.7cm}
\caption{Comparison of reported $(g-r)$ CMR gradients from
other X-ray selected cluster surveys with the present work.
No trend is apparent with mean $L_X$ (left), but this may reflect
the nature of the individual clusters in each sample.
Conversely, these works strongly suggest 
a steeper radial modal colour gradient with redshift (right).
}
        \label{fig:compare}
\end{figure*}

In Table~\ref{tab:grads} we give the colour gradient for the inner regions
of the composite cluster to the limit where the gradient becomes zero 
for all of the SDSS colours.  This is the most comprehensive 
reporting of modal colour variations across clusters to date that we are
aware of.  Hence to
compare this with other studies, we 
must restrict ourselves to more commonly used colours -- e.g.\
our reported $(g-r)$ colour gradient.  
In Fig.~\ref{fig:compare} we display the radial $(g-r)$ gradient 
with the results reported by 
Abraham et al.\ (1996), Terlevich et al.\ (2001), 
Pimbblet et al.\ (2002), and Wake et al.\ (2005)\footnote{We transform
the colours reported in these studies to $(g-r)$ colours
through the transforms presented by Smith et al.\ (2002).} 
together with the gradient as a function of 
mean $L_X$ for each of those studies. 

\begin{table}
\begin{center}
\caption{Modal colour gradient for the various SDSS colours
as a function of projected radius and local galaxy density.
\hfil}
\begin{tabular}{ll}
\noalign{\medskip}
\hline
Radial Gradient & Value  \\
\hline
$d(u-g)/d log(r_p)$ & $-0.069 \pm 0.007$ \\
$d(u-r)/d log(r_p)$ & $-0.099 \pm 0.008$ \\
$d(u-i)/d log(r_p)$ & $-0.130 \pm 0.020$ \\
$d(u-z)/d log(r_p)$ & $-0.130 \pm 0.010$ \\
$d(g-r)/d log(r_p)$ & $-0.031 \pm 0.003$ \\
$d(g-i)/d log(r_p)$ & $-0.049 \pm 0.004$ \\
$d(g-z)/d log(r_p)$ & $-0.073 \pm 0.006$ \\
$d(r-i)/d log(r_p)$ & $-0.018 \pm 0.002$ \\
$d(r-z)/d log(r_p)$ & $-0.035 \pm 0.004$ \\
$d(i-z)/d log(r_p)$ & $-0.015 \pm 0.002$ \\
\hline
Density Gradient & Value \\ 
\hline
$d(u-g)/d log(\Sigma)$ & $0.025 \pm 0.004$ \\
$d(u-r)/d log(\Sigma)$ & $0.041 \pm 0.004$ \\
$d(u-i)/d log(\Sigma)$ & $0.043 \pm 0.006$ \\
$d(u-z)/d log(\Sigma)$ & $0.048 \pm 0.007$ \\
$d(g-r)/d log(\Sigma)$ & $0.012 \pm 0.002$ \\
$d(g-i)/d log(\Sigma)$ & $0.020 \pm 0.003$ \\
$d(g-z)/d log(\Sigma)$ & $0.024 \pm 0.005$ \\
$d(r-i)/d log(\Sigma)$ & $0.007 \pm 0.001$ \\
$d(r-z)/d log(\Sigma)$ & $0.017 \pm 0.002$ \\
$d(i-z)/d log(\Sigma)$ & $0.009 \pm 0.001$ \\
\hline
\end{tabular}
  \label{tab:grads}
\end{center}
\end{table}

There appears to be no correlation of X-ray luminosity -- and
therefore cluster mass by implication -- for the radial colour
gradients. However, Abraham et al.\ (1996) and Terlevich et al.\ (2001)
are only single-case studies (Abell~2370 and Coma respectively).  
Therefore the gradients reported in those works may 
reflect the individual accretion history of those clusters rather than
the more general results presented here and by Pimbblet et al.\ (2002)
and Wake et al.\ (2005).  This is supported by the results of Stott et al.\ (2009)
who find the CMR slope of Coma is very unusual compared to other clusters.
Moreover, the range of $L_X$ probed by Wake et al.\ (2005) means that that
point should also be regarded as tentative.  
For our study and Pimbblet et al.\ (2002), the cluster samples are intentionally
limited in $L_X$ and we therefore hypothesize that there may yet be a relationship
between colour gradient and $L_X$ such that the gradient becomes steeper
with increasing $L_X$ values.  
A targeted study of a carefully constructed sample of low $L_X$ clusters 
is therefore urgently needed to attempt to falsify this hypothesis.

Applying the same logic to the redshift evolution, we contend that the
radial variation of the modal CMR colour gets increasingly strong at higher redshift.
That said, we are not necessarily comparing like clusters with each other
over that redshift range -- the progenitors of our sample are most certainly not
going to be higher $L_X$ clusters at higher redshifts.  
Given the median difference in $L_X$ from our sample to (e.g.) LARCS, we
estimate that the difference in total mass would be a factor of $\sim$ 2--3 
(Popesso et al.\ 2005; see also Stott et al.\ 2009).  Since we do not
expect significant variation in these cluster's luminosity functions 
over such a mass range (cf.\ De Propris et al.\ 1999), we would expect 
that any variation in cluster properties are reflective of evolutionary
change over the redshift range.
Further, we note that the range of $L_X$
probed by Wake et al.\ (2005) does include some clusters in our own $L_X$ 
interval, which adds weight to the above argument for redshift evolution.  
This trend would also
be straight-forward to interpret: the red cores of clusters establish themselves
at early times and gain mass through accreting bluer `field' galaxies which
progressively redden over time (cf.\ De Lucia et al.\ 2007; Kodama \& Bower 2001;
Mart{\'{\i}}nez, Coenda \& Muriel 2010).
The colour gradient therefore becomes shallower as the blue galaxies 
transform to redder galaxies at higher radii from cluster centres at a fixed mass
and reflects the hierarchical build-up of the red sequence over time.

\section{Summary}
In this work, we have assembled a sample of intermediate X-ray 
luminosity galaxy clusters from SDSS photometry and spectroscopy.
Our principal results are as follows:\\

\noindent $\bullet$ 
Strong colour-magnitude red sequences exist for our $L_X$ range
in all SDSS colours probed.  As with higher $L_X$ studies, these
CMRs can be accessed to a large radius from the cluster centres, out in
to the surrounding filamentary regime.\\

\noindent $\bullet$
The clusters exhibit a modal colour gradient for the red sequence galaxies 
in projected radius and local galaxy density whereby red sequence galaxies
at the cluster outskirts are systematically bluer than those in the core
for all SDSS colours.
These gradients are interpreted as the galaxies at the cluster outskirts
having younger luminosity-weighted ages.\\

\noindent $\bullet$
Our results agree with earlier measurements of colour gradients and extend
them to comprehensively cover optical and near infra-red colours.  
They suggest that there may be a relationship between 
redshift and CMR colour gradient, but further studies 
using carefully constructed cluster samples are needed to verify
this hypothesis and any potential relationship between $L_X$ and
colour gradient.

\section*{Acknowledgements}
PCJ and KAP contributed equally to this manuscript.

PCJ acknowledges receipt of an Australian Postgraduate Award. 
KAP acknowledges partial support from the Australian Research Council
and a Monash University internal grant scheme.

We thank Warrick Couch and Michael Drinkwater for helpful feedback on the content
of this paper which is derived from PCJ's Honours Thesis, awarded by U.Queensland.
Further, we gratefully thank the Astronomical Society of Australia for the 
award of the Bok Prize (2009) to PCJ for said thesis.

We also thank the anonymous referee for her/his 
constructive comments that have improved the work reported here.

Funding for the SDSS and SDSS-II has been provided by the 
Alfred P.\ Sloan Foundation, the Participating Institutions, 
the National Science Foundation, the U.S. Department of Energy, 
the National Aeronautics and Space Administration, the Japanese 
Monbukagakusho, the Max Planck Society, and the Higher Education 
Funding Council for England. The SDSS Web Site is http://www.sdss.org/.

The SDSS is managed by the Astrophysical Research Consortium for the 
Participating Institutions. The Participating Institutions are the 
American Museum of Natural History, Astrophysical Institute Potsdam, 
University of Basel, University of Cambridge, Case Western Reserve 
University, University of Chicago, Drexel University, Fermilab, the 
Institute for Advanced Study, the Japan Participation Group, Johns 
Hopkins University, the Joint Institute for Nuclear Astrophysics, the 
Kavli Institute for Particle Astrophysics and Cosmology, the Korean 
Scientist Group, the Chinese Academy of Sciences (LAMOST), Los Alamos 
National Laboratory, the Max-Planck-Institute for Astronomy (MPIA), the 
Max-Planck-Institute for Astrophysics (MPA), New Mexico State University, 
Ohio State University, University of Pittsburgh, University of Portsmouth, 
Princeton University, the United States Naval Observatory, and the University of Washington.

This research has made use of the NASA/IPAC Extragalactic Database (NED) which is 
operated by the Jet Propulsion Laboratory, California Institute of Technology, 
under contract with the National Aeronautics and Space Administration.

\medskip
\section*{Appendix~A: Discordant points on the colour-magnitude planes}

\begin{figure*}
\centerline{\psfig{file=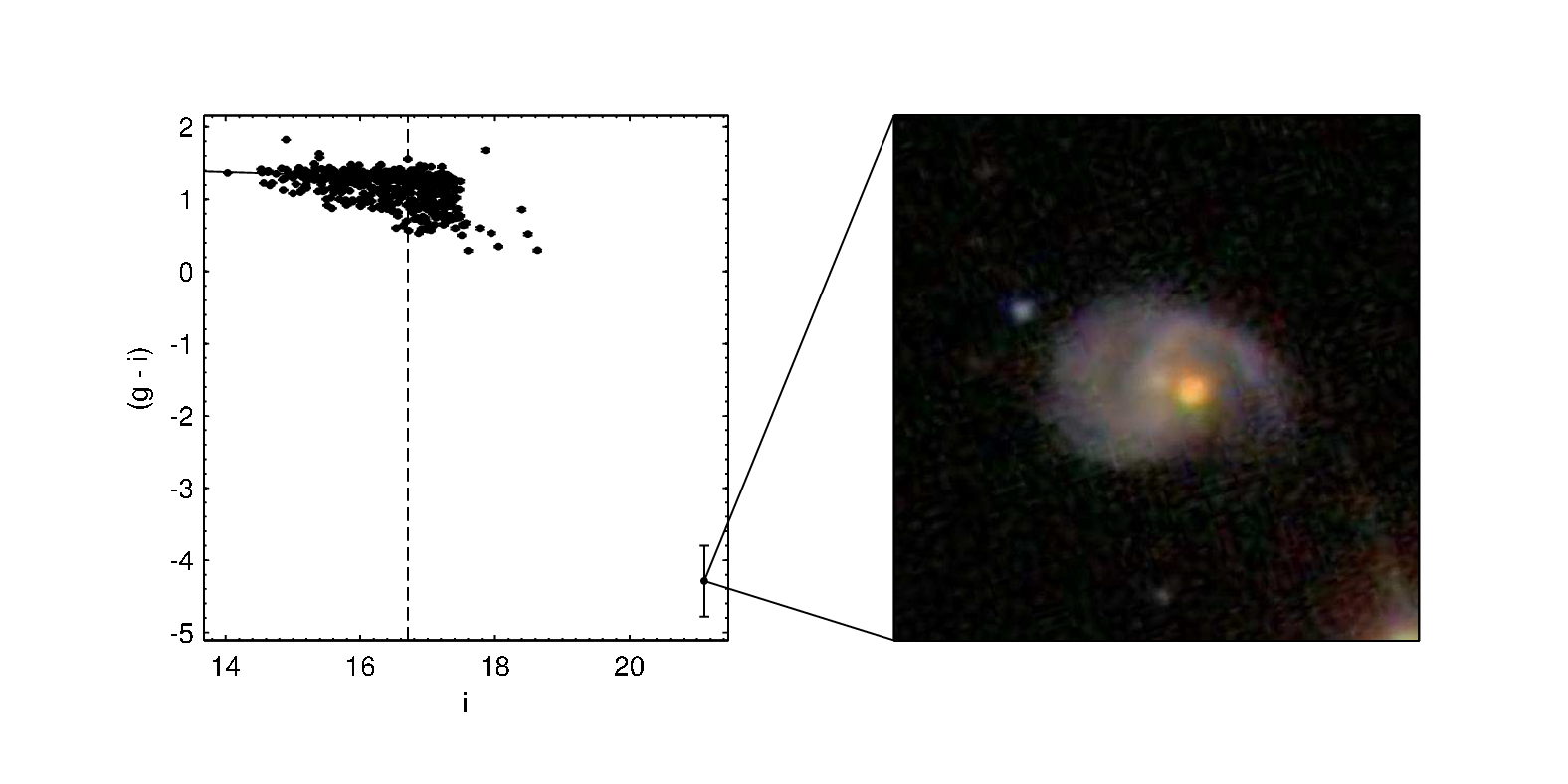,angle=0,width=6.75in}}
\vspace*{-0.7cm}
\caption{Discordant point on the (g-i) versus i colour-magnitude
diagram for Abell 1767. A blue spiral galaxy in the line-of-sight
of the target red elliptical galaxy confuses the photometric
measurements, causing the i band to be much fainter than the other bands.
}
        \label{fig:discord1}
\end{figure*}

\begin{figure*}
\centerline{\psfig{file=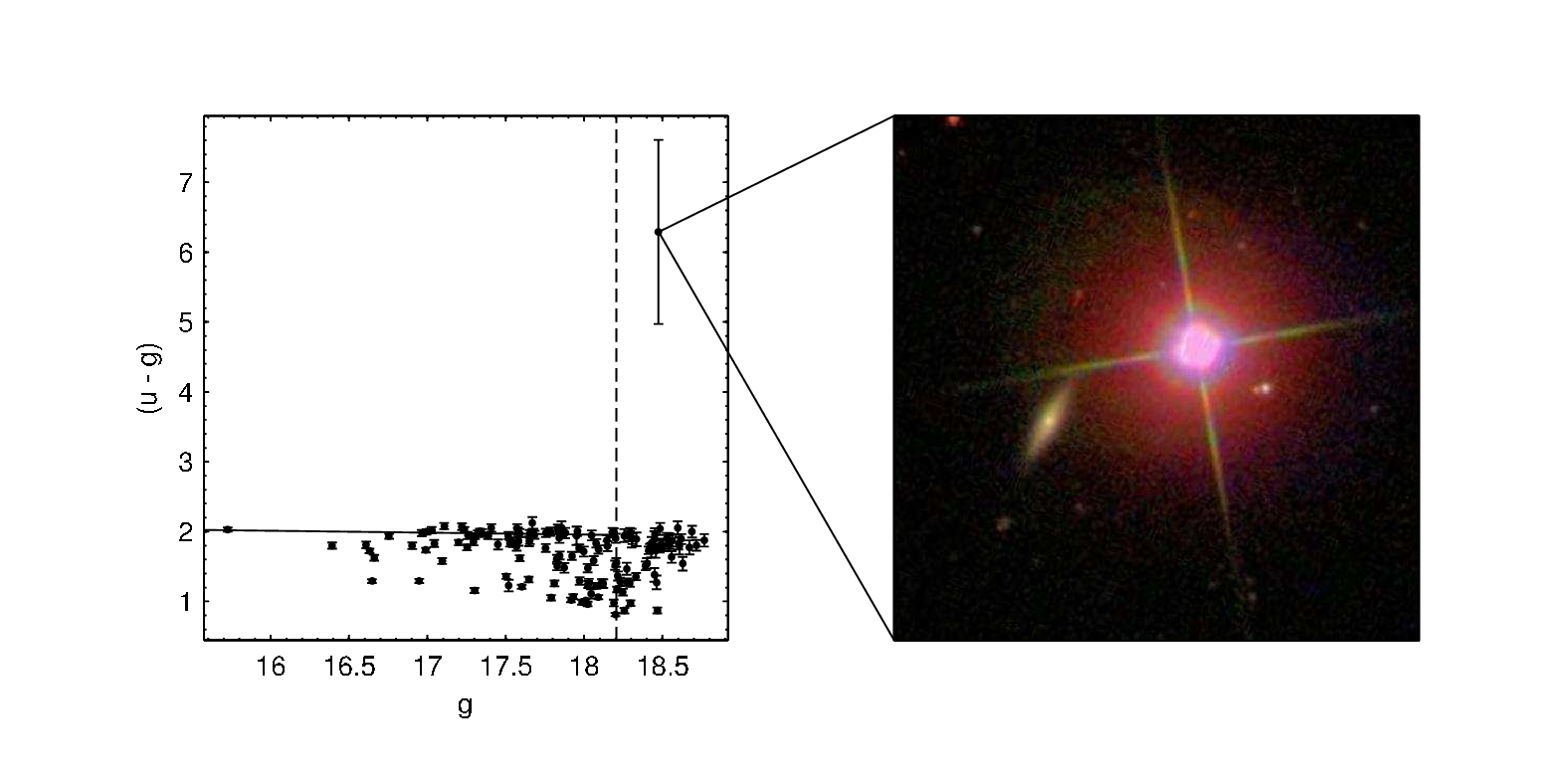,angle=0,width=6.75in}}
\vspace*{-0.7cm}
\caption{Discordant point on the (u-i) versus u colour-magnitude
diagram for Abell 971. A near-by saturated star affects the
photometry of the target galaxy.
}
        \label{fig:discord2}
\end{figure*}
   
\begin{figure*}
\centerline{\psfig{file=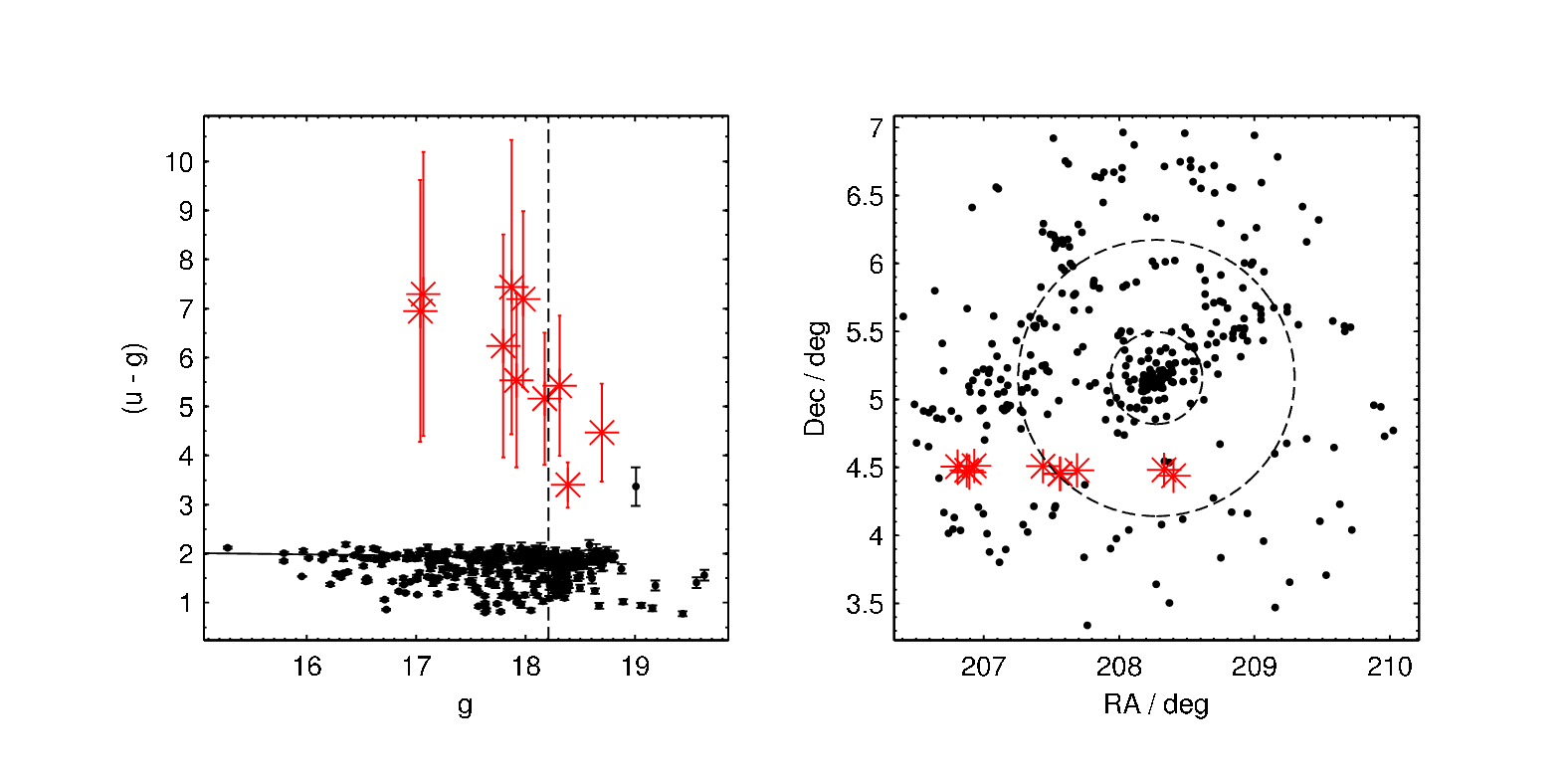,angle=0,width=6.75in}}
\vspace*{-0.7cm}
\caption{A series of discordant points in Abell~1809 that appear to
congregate around 4.5 degrees of declination on the RA-Dec diagram.
No bad pixel rows or nearby
sources of photometric contamination have been found for these
points.  The tightness in Dec is suggestive of an observational
artifact.
}
        \label{fig:discord3}
\end{figure*}

In creating the colour-magnitude diagrams for our cluster sample (Fig.~\ref{fig:cmrs}), we
located a number of discordant galaxies whose position on the colour-magnitude
plane was very much away from the bulk of the population (i.e.\ 
much redder than the red sequence or much bluer than the blue cloud).

The majority of the discordant points appear to be caused by photometric 
saturation from other objects nearby to the target galaxy. The most common 
source of colour bias is due to superpositions of line-of-sight galaxies. 
A typical example is the discordant point on the (g-i) versus i 
colour-magnitude diagram for Abell 1767 (Fig.~\ref{fig:discord1}). 
The same discordant point also lies far below the blue cloud in the 
(u-i) versus i, and (r-i) versus i colour-magnitude diagrams and far 
above the red sequence in the (i-z) versus z plane.

Photometric saturation from a bright near-by stars also affects the
photometry of a number of galaxies.  Fig.~\ref{fig:discord2} shows
one such point on the (u-g) versus g colour-magnitude diagram for Abell 971. 

Dust reddening appears to affect other galaxies, as determined from a visual
inspection of their image (i.e.\ obvious dust lanes).  
However, a unique set of galaxies in Abell~1809 
appears to possess very red colours (Fig.~\ref{fig:discord3}).  We were unable
to determine the cause of these galaxies anomalous colours.

In total, the amount of contamination in the photometric galaxy catalogue is very small.
Of the 10,529 galaxies, only 46 galaxies (or 0.4 per cent) have colours which are 
atypical ($3\sigma$ above or $5\sigma$ below 
the mean galaxian colour of their host cluster). 
Of these 46 discordant points, 
13 are due to line-of-sight galaxy effects (Fig.~\ref{fig:discord1}), 
11 are due to saturation from nearby stars (Fig.~\ref{fig:discord2}),
and 
12 are suspected to be due to dust reddening, or the situation outlined
in Fig.~\ref{fig:discord3}.
The majority of the remaining discordant points appear to be genuinely blue galaxies. 
Of the 7 genuinely blue galaxy, 
4 are spectroscopically confirmed low-redshift QSOs. 
There were also 2 genuinely red galaxies and 1 galaxy which was bifurcated 
at the edge of a stripe.

Given we have $\sim$10,000 objects in our sample, we may have expected $\sim100$
to have significantly unusual colours (by our above definition).  
Our 46 outliers is slightly fewer than this expectation value, but is 
not worrisome, particularly as we look for $>5\sigma$ bluer 
than the mean galaxy colour to avoid the blue cloud.


\begin{thebibliography}{}

\bibitem[\protect\citeauthoryear{Abraham et 
al.}{1996}]{1996ApJ...471..694A} Abraham R.~G., et al., 1996, ApJ, 471, 694 

\bibitem[\protect\citeauthoryear{Adelman-McCarthy et 
al.}{2008}]{2008ApJS..175..297A} Adelman-McCarthy J.~K., et al., 2008, 
ApJS, 175, 297 

\bibitem[\protect\citeauthoryear{Andreon et 
al.}{2004}]{2004MNRAS.353..353A} Andreon S., Willis J., Quintana H., 
Valtchanov I., Pierre M., Pacaud F., 2004, MNRAS, 353, 353 

\bibitem[]{1994AJ....108.2348A} Ashman, K. M., Bird, C. M., \& Zepf, S. E., 1994, AJ, 108, 2348

\bibitem[\protect\citeauthoryear{Babul et al.}{2002}]{2002MNRAS.330..329B} 
Babul A., Balogh M.~L., Lewis G.~F., Poole G.~B., 2002, MNRAS, 330, 329 

\bibitem[\protect\citeauthoryear{Balogh et al.}{1999}]{1999ApJ...527...54B} 
Balogh M.~L., Morris S.~L., Yee H.~K.~C., Carlberg R.~G., Ellingson E., 
1999, ApJ, 527, 54 

\bibitem[\protect\citeauthoryear{Balogh, Navarro, 
\& Morris}{2000}]{2000ApJ...540..113B} Balogh M.~L., Navarro J.~F., Morris S.~L., 2000, ApJ, 540, 113 

\bibitem[\protect\citeauthoryear{Bautz 
\& Morgan}{1970}]{1970ApJ...162L.149B} Bautz L.~P., Morgan W.~W., 1970, ApJ, 162, L149 

\bibitem[\protect\citeauthoryear{Bernardi et 
al.}{2006}]{2006AJ....131.1288B} Bernardi M., Nichol R.~C., Sheth R.~K., 
Miller C.~J., Brinkmann J., 2006, AJ, 131, 1288 

\bibitem[\protect\citeauthoryear{Blanton et 
al.}{2003}]{2003AJ....125.2276B} Blanton M.~R., Lin H., Lupton R.~H., Maley 
F.~M., Young N., Zehavi I., Loveday J., 2003, AJ, 125, 2276 

\bibitem[\protect\citeauthoryear{Bower, Lucey, 
\& Ellis}{1992}]{1992MNRAS.254..589B} Bower R.~G., Lucey J.~R., Ellis R.~S., 1992, MNRAS, 254, 589 

\bibitem[\protect\citeauthoryear{Burgett et 
al.}{2004}]{2004MNRAS.352..605B} Burgett W.~S., et al., 2004, MNRAS, 352, 
605 

\bibitem[\protect\citeauthoryear{Butcher 
\& Oemler}{1984}]{1984ApJ...285..426B} Butcher H., Oemler A., Jr., 1984, ApJ, 285, 426 

\bibitem[\protect\citeauthoryear{Byrd 
\& Valtonen}{1990}]{1990ApJ...350...89B} Byrd G., Valtonen M., 1990, ApJ, 350, 89 

\bibitem[\protect\citeauthoryear{Carlberg, Yee, 
\& Ellingson}{1997}]{1997ApJ...478..462C} Carlberg R.~G., Yee H.~K.~C., Ellingson E., 1997, ApJ, 478, 462 

\bibitem[\protect\citeauthoryear{Cortese 
\& Hughes}{2009}]{2009MNRAS.400.1225C} Cortese L., Hughes T.~M., 2009, MNRAS, 400, 1225 

\bibitem[\protect\citeauthoryear{Dahl{\'e}n et 
al.}{2004}]{2004MNRAS.350..253D} Dahl{\'e}n T., Fransson C., {\"O}stlin G., 
N{\"a}slund M., 2004, MNRAS, 350, 253 

\bibitem[\protect\citeauthoryear{De Lucia et 
al.}{2004}]{2004ApJ...610L..77D} De Lucia G., et al., 2004, ApJ, 610, L77 

\bibitem[\protect\citeauthoryear{De Lucia et 
al.}{2007}]{2007MNRAS.374..809D} De Lucia G., et al., 2007, MNRAS, 374, 809 

\bibitem[\protect\citeauthoryear{Diaferio et 
al.}{2001}]{2001MNRAS.323..999D} Diaferio A., Kauffmann G., Balogh M.~L., 
White S.~D.~M., Schade D., Ellingson E., 2001, MNRAS, 323, 999 

\bibitem[\protect\citeauthoryear{Dressler}{1980}]{1980ApJ...236..351D} 
Dressler A., 1980, ApJ, 236, 351 

\bibitem[]{1988AJ.....95..985D} Dressler, A., \& Shectman, S. A., 1988, AJ 95, 985  

\bibitem[\protect\citeauthoryear{Dressler et 
al.}{1997}]{1997ApJ...490..577D} Dressler A., et al., 1997, ApJ, 490, 577 

\bibitem[\protect\citeauthoryear{Ebeling et 
al.}{1996}]{1996MNRAS.281..799E} Ebeling H., Voges W., Bohringer H., Edge 
A.~C., Huchra J.~P., Briel U.~G., 1996, MNRAS, 281, 799 

\bibitem[\protect\citeauthoryear{Ebeling et 
al.}{1998}]{1998MNRAS.301..881E} Ebeling H., Edge A.~C., Bohringer H., 
Allen S.~W., Crawford C.~S., Fabian A.~C., Voges W., Huchra J.~P., 1998, 
MNRAS, 301, 881 

\bibitem[\protect\citeauthoryear{Ebeling et 
al.}{2000}]{2000MNRAS.318..333E} Ebeling H., Edge A.~C., Allen S.~W., 
Crawford C.~S., Fabian A.~C., Huchra J.~P., 2000, MNRAS, 318, 333 

\bibitem[\protect\citeauthoryear{Ebeling, Edge, 
\& Henry}{2001}]{2001ApJ...553..668E} Ebeling H., Edge A.~C., Henry J.~P., 2001, ApJ, 553, 668 

\bibitem[\protect\citeauthoryear{Fairley et 
al.}{2002}]{2002MNRAS.330..755F} Fairley B.~W., Jones L.~R., Wake D.~A., 
Collins C.~A., Burke D.~J., Nichol R.~C., Romer A.~K., 2002, MNRAS, 330, 
755 

\bibitem[\protect\citeauthoryear{Fasano et al.}{2000}]{2000ApJ...542..673F} 
Fasano G., Poggianti B.~M., Couch W.~J., Bettoni D., Kj{\ae}rgaard P., 
Moles M., 2000, ApJ, 542, 673 

\bibitem[\protect\citeauthoryear{Fasano et 
al.}{2006}]{2006A&A...445..805F} Fasano G., et al., 2006, A\&A, 445, 805 

\bibitem[\protect\citeauthoryear{Girardi et 
al.}{2003}]{2003A&A...410..461G} Girardi M., Mardirossian F., Marinoni C., Mezzetti M., 
Rigoni E., 2003, A\&A, 410, 461 

\bibitem[\protect\citeauthoryear{G{\'o}mez et 
al.}{2003}]{2003ApJ...584..210G} G{\'o}mez P.~L., et al., 2003, ApJ, 584, 
210 

\bibitem[\protect\citeauthoryear{Gunn 
\& Gott}{1972}]{1972ApJ...176....1G} Gunn J.~E., Gott J.~R., III, 1972, ApJ, 176, 1 

\bibitem[\protect\citeauthoryear{Haas, Schaye, 
\& Jeeson-Daniel}{2011}]{2011MNRAS.tmp.1812H} Haas M.~R., Schaye J., Jeeson-Daniel A., 2011, MNRAS, 1812 

\bibitem[\protect\citeauthoryear{Haines et 
al.}{2004}]{2004A&A...425..783H} Haines C.~P., 
Mercurio A., Merluzzi P., La Barbera F., Massarotti M., Busarello G., Girardi M., 2004, A\&A, 425, 783 

\bibitem[\protect\citeauthoryear{Haines et al.}{2006}]{2006MNRAS.371...55H} 
Haines C.~P., Merluzzi P., Mercurio A., Gargiulo A., Krusanova N., 
Busarello G., La Barbera F., Capaccioli M., 2006, MNRAS, 371, 55 

\bibitem[Hogg et al.(2002)]{2002astro.ph.10394H} Hogg D.~W., Baldry 
I.~K., Blanton M.~R., Eisenstein D.~J., 2002, arXiv:astro-ph/0210394 

\bibitem[\protect\citeauthoryear{Hogg et al.}{2004}]{2004ApJ...601L..29H} 
Hogg D.~W., et al., 2004, ApJ, 601, L29 

\bibitem[\protect\citeauthoryear{Huertas-Company et 
al.}{2009}]{2009A&A...505...83H} Huertas-Company M., Foex G., Soucail G., Pell{\'o} R., 2009, A\&A, 505, 83 

\bibitem[\protect\citeauthoryear{Jim{\'e}nez et 
al.}{2011}]{2011MNRAS.417..785J} Jim{\'e}nez N., Cora S.~A., Bassino L.~P., 
Tecce T.~E., Smith Castelli A.~V., 2011, MNRAS, 417, 785 

\bibitem[\protect\citeauthoryear{Kodama 
\& Arimoto}{1997}]{1997A&A...320...41K} Kodama T., Arimoto N., 1997, A\&A, 320, 41 

\bibitem[\protect\citeauthoryear{Kodama, Bower, 
\& Bell}{1999}]{1999MNRAS.306..561K} Kodama T., Bower R.~G., Bell E.~F., 1999, MNRAS, 306, 561 

\bibitem[\protect\citeauthoryear{Kodama 
\& Bower}{2001}]{2001MNRAS.321...18K} Kodama T., Bower R.~G., 2001, MNRAS, 321, 18 

\bibitem[\protect\citeauthoryear{Lacey 
\& Cole}{1993}]{1993MNRAS.262..627L} Lacey C., Cole S., 1993, MNRAS, 262, 627 

\bibitem[\protect\citeauthoryear{Larson, Tinsley, 
\& Caldwell}{1980}]{1980ApJ...237..692L} Larson R.~B., Tinsley B.~M., Caldwell C.~N., 1980, ApJ, 237, 692 

\bibitem[\protect\citeauthoryear{Lewis et al.}{2002}]{2002MNRAS.334..673L} 
Lewis I., et al., 2002, MNRAS, 334, 673 

\bibitem[\protect\citeauthoryear{Ma et al.}{2010}]{2010MNRAS.406..121M} Ma 
C.-J., Ebeling H., Marshall P., Schrabback T., 2010, MNRAS, 406, 121 

\bibitem[\protect\citeauthoryear{Mahajan 
\& Raychaudhury}{2009}]{2009MNRAS.400..687M} Mahajan S., Raychaudhury S., 2009, MNRAS, 400, 687 

\bibitem[\protect\citeauthoryear{Mart{\'{\i}}nez, Coenda, 
\& Muriel}{2010}]{2010MNRAS.403..748M} Mart{\'{\i}}nez H.~J., Coenda V., Muriel H., 2010, MNRAS, 403, 748 

\bibitem[\protect\citeauthoryear{McIntosh, Rix, 
\& Caldwell}{2004}]{2004ApJ...610..161M} McIntosh D.~H., Rix H.-W., Caldwell N., 2004, ApJ, 610, 161 

\bibitem[\protect\citeauthoryear{Mei et al.}{2009}]{2009ApJ...690...42M} 
Mei S., et al., 2009, ApJ, 690, 42 

\bibitem[\protect\citeauthoryear{Mobasher et 
al.}{2003}]{2003ApJ...587..605M} Mobasher B., et al., 2003, ApJ, 587, 605 

\bibitem[\protect\citeauthoryear{Moore et al.}{1996}]{1996Natur.379..613M} 
Moore B., Katz N., Lake G., Dressler A., Oemler A., 1996, Natur, 379, 613 

\bibitem[\protect\citeauthoryear{Muldrew et 
al.}{2011}]{2011MNRAS.tmp.1915M} Muldrew S.~I., et al., 2011, MNRAS, in press

\bibitem[\protect\citeauthoryear{Pimbblet, Drinkwater, 
\& Hawkrigg}{2004}]{2004MNRAS.354L..61P} Pimbblet K.~A., 
Drinkwater M.~J., Hawkrigg M.~C., 2004, MNRAS, 354, L61 

\bibitem[\protect\citeauthoryear{Pimbblet et 
al.}{2001}]{2001MNRAS.327..588P} Pimbblet K.~A., Smail I., Edge A.~C., 
Couch W.~J., O'Hely E., Zabludoff A.~I., 2001, MNRAS, 327, 588 

\bibitem[\protect\citeauthoryear{Pimbblet et 
al.}{2002}]{2002MNRAS.331..333P} Pimbblet K.~A., Smail I., Kodama T., Couch 
W.~J., Edge A.~C., Zabludoff A.~I., O'Hely E., 2002, MNRAS, 331, 333 

\bibitem[\protect\citeauthoryear{Pimbblet et 
al.}{2006}]{2006MNRAS.366..645P} Pimbblet K.~A., Smail I., Edge A.~C., 
O'Hely E., Couch W.~J., Zabludoff A.~I., 2006, MNRAS, 366, 645 

\bibitem[\protect\citeauthoryear{Pimbblet et 
al.}{2011}]{2011MNRAS.410.1837P} Pimbblet K.~A., Andernach H., Fishlock 
C.~K., Roseboom I.~G., Owers M.~S., 2011, MNRAS, 410, 1837 

\bibitem[\protect\citeauthoryear{Poggianti et 
al.}{2004}]{2004ApJ...601..197P} Poggianti B.~M., Bridges T.~J., Komiyama 
Y., Yagi M., Carter D., Mobasher B., Okamura S., Kashikawa N., 2004, ApJ, 
601, 197 

\bibitem[\protect\citeauthoryear{Poggianti et 
al.}{2009}]{2009ApJ...697L.137P} Poggianti B.~M., et al., 2009, ApJ, 697, 
L137 

\bibitem[\protect\citeauthoryear{Popesso et 
al.}{2005}]{2005A&A...433..415P} Popesso P., 
B{\"o}hringer H., Romaniello M., Voges W., 2005, A\&A, 433, 415 

\bibitem[\protect\citeauthoryear{Press et al.}{1992}]{1992nrfa.book.....P} 
Press W.~H., Teukolsky S.~A., Vetterling W.~T., Flannery B.~P., 1992, 
Numerical Recipies, Cambridge University Press, Cambridge

\bibitem[\protect\citeauthoryear{Quilis, Moore, 
\& Bower}{2000}]{2000Sci...288.1617Q} Quilis V., Moore B., Bower R., 2000, Sci, 288, 1617 

\bibitem[\protect\citeauthoryear{Sato 
\& Martin}{2006}]{2006ApJ...647..946S} Sato T., Martin C.~L., 2006, ApJ, 647, 946 

\bibitem[\protect\citeauthoryear{Schechter}{1976}]{1976ApJ...203..297S} 
Schechter P., 1976, ApJ, 203, 297 

\bibitem[\protect\citeauthoryear{Smith et al.}{2002}]{2002AJ....123.2121S} 
Smith J.~A., et al., 2002, AJ, 123, 2121 

\bibitem[\protect\citeauthoryear{Smith et al.}{2006}]{2006MNRAS.369.1419S} 
Smith R.~J., Hudson M.~J., Lucey J.~R., Nelan J.~E., Wegner G.~A., 2006, 
MNRAS, 369, 1419 

\bibitem[\protect\citeauthoryear{Spergel et 
al.}{2007}]{2007ApJS..170..377S} Spergel D.~N., et al., 2007, ApJS, 170, 
377 

\bibitem[\protect\citeauthoryear{Stott et al.}{2007}]{2007ApJ...661...95S} 
Stott J.~P., Smail I., Edge A.~C., Ebeling H., Smith G.~P., Kneib J.-P., 
Pimbblet K.~A., 2007, ApJ, 661, 95 

\bibitem[\protect\citeauthoryear{Stott et al.}{2009}]{2009MNRAS.394.2098S} 
Stott J.~P., Pimbblet K.~A., Edge A.~C., Smith G.~P., Wardlow J.~L., 2009, 
MNRAS, 394, 2098 

\bibitem[\protect\citeauthoryear{Strauss et 
al.}{2002}]{2002AJ....124.1810S} Strauss M.~A., et al., 2002, AJ, 124, 1810 

\bibitem[\protect\citeauthoryear{Tanaka et al.}{2005}]{2005MNRAS.362..268T} 
Tanaka M., Kodama T., Arimoto N., Okamura S., Umetsu K., Shimasaku K., 
Tanaka I., Yamada T., 2005, MNRAS, 362, 268 

\bibitem[\protect\citeauthoryear{Terlevich, Caldwell, 
\& Bower}{2001}]{2001MNRAS.326.1547T} Terlevich A.~I., Caldwell N., Bower R.~G., 2001, MNRAS, 326, 1547 

\bibitem[\protect\citeauthoryear{Thompson 
\& Gregory}{1993}]{1993AJ....106.2197T} Thompson L.~A., Gregory S.~A., 1993, AJ, 106, 2197 

\bibitem[\protect\citeauthoryear{Tran et al.}{2005}]{2005ApJ...619..134T} 
Tran K.-V.~H., van Dokkum P., Illingworth G.~D., Kelson D., Gonzalez A., 
Franx M., 2005, ApJ, 619, 134 

\bibitem[\protect\citeauthoryear{Valentinuzzi et 
al.}{2011}]{2011A&A...536A..34V} Valentinuzzi T., et al., 2011, A\&A, 536, A34 

\bibitem[\protect\citeauthoryear{Visvanathan 
\& Sandage}{1977}]{1977ApJ...216..214V} Visvanathan N., Sandage A., 1977, ApJ, 216, 214 

\bibitem[\protect\citeauthoryear{Wake et al.}{2005}]{2005ApJ...627..186W} 
Wake D.~A., Collins C.~A., Nichol R.~C., Jones L.~R., Burke D.~J., 2005, 
ApJ, 627, 186 

\bibitem[\protect\citeauthoryear{Yahil 
\& Vidal}{1977}]{1977ApJ...214..347Y} Yahil A., Vidal N.~V., 1977, ApJ, 214, 347 

\bibitem[\protect\citeauthoryear{Yee, Ellingson, 
\& Carlberg}{1996}]{1996ApJS..102..269Y} Yee H.~K.~C., Ellingson E., Carlberg R.~G., 1996, ApJS, 102, 269 

\bibitem[\protect\citeauthoryear{York et al.}{2000}]{2000AJ....120.1579Y} 
York D.~G., et al., 2000, AJ, 120, 1579 

\end{thebibliography}
\end{document}